\documentclass[reprint, onecolumn, notitlepage, longbibliography, aps, nofootinbib, showpacs, superscriptaddress, prb, preprintnumbers]{revtex4-2}
\pdfoutput=1
\usepackage[utf8]{inputenc}
\usepackage{amsmath,amssymb,graphicx}
\usepackage{color}
\usepackage{caption, subcaption}
\usepackage{braket}

\newcommand{\DO}{\Delta_{\mathcal{O}}}
\newcommand{\Dpsi}{\Delta_{\mathcal{\psi}}}

\def\bea{\begin{equation}\begin{aligned}}
\def\eea{\end{aligned}\end{equation}}
\def \tr {\text{tr}}

\graphicspath{{../Manuscript/img/}}

\begin{document}
    
    \preprint{UTTG--06--2019}
    
    \title{The Speed of Quantum Information Spreading in Chaotic Systems}
    
    \author{Josiah Couch}
    \affiliation{Weinberg Theory Group, Department of Physics, The University of Texas at Austin,
        Austin, TX 78712, USA}
    \author{Stefan Eccles}
    \affiliation{Weinberg Theory Group, Department of Physics, The University of Texas at Austin,
        Austin, TX 78712, USA}
    \author{Phuc Nguyen}
    \affiliation{Maryland Center for Fundamental Physics, University of Maryland, College Park, MD 20742, USA}
    \author{Brian Swingle}
    \affiliation{Maryland Center for Fundamental Physics, University of Maryland, College Park, MD 20742, USA}
    \affiliation{Condensed Matter Theory Center, University of Maryland, College Park, MD 20742, USA}
    \affiliation{Joint Institute for Quantum Information and Computer Science, University of Maryland, College Park, MD 20742, USA}
    \author{Shenglong Xu}
    \affiliation{Condensed Matter Theory Center, University of Maryland, College Park, MD 20742, USA}
    \affiliation{Department of Physics \& Astronomy, Texas A\&M University, College Station, Texas 77843, USA}

\begin{abstract}

We present a general theory of quantum information propagation in chaotic quantum many-body systems. The generic expectation in such systems is that quantum information does not propagate in localized form; instead, it tends to spread out and scramble into a form that is inaccessible to local measurements. To characterize this spreading, we define an information speed via a quench-type experiment and derive a general formula for it as a function of the entanglement density of the initial state. As the entanglement density varies from zero to one, the information speed varies from the entanglement speed to the butterfly speed. We verify that the formula holds both for a quantum chaotic spin chain and in field theories with an AdS/CFT gravity dual. For the second case, we study in detail the dynamics of entanglement in two-sided Vaidya-AdS-Reissner-Nordstrom black branes. We also show that, with an appropriate decoding process, quantum information can be construed as moving at the information speed, and, in the case of AdS/CFT, we show that a locally detectable signal propagates at the information speed in a spatially local variant of the traversable wormhole setup.

\end{abstract}
    
\maketitle

\tableofcontents

\section{Introduction and results}


Quantum information provides a unifying language that bridges the subjects of quantum many-body physics, quantum field theory, and quantum gravity. For example, in many-body physics, entanglement plays a crucial role in the classification of quantum phases of matter~\cite{Kitaev2006,Li2008,Vedral2008} and leads to powerful numerical tools \cite{Vidal2007,Schollwock2011}. In AdS/CFT \cite{Maldacena:1997re, Witten:1998qj, Gubser:1998bc}, it underlies the holographic emergence of spacetime~\cite{Ryu:2006bv,VanRaamsdonk_2010_Entanglement_Spacetime}. More generally, ideas and tools related to computational complexity \cite{Susskind:2014moa, Stanford:2014jda, Brown:2015lvg, Brown:2015bva}, entanglement \cite{Ryu:2006ef, Ryu:2006bv, Faulkner:2013ica, Faulkner:2017tkh}, quantum error correction \cite{Almheiri:2014lwa, Harlow:2016vwg}, and much else have led to new insights into all of the above areas of physics.


Our focus here is on the dynamics of quantum information in chaotic systems. The effect of interest is the scrambling of initially simple information into a form that is inaccessible to local measurements~\cite{Hayden:2007cs,Sekiro_2008_Scrambling,Brown_2012_Scrambling}. For example, in a chaotic system, two initial states with the same average energy but differing in some non-conserved local observable will rapidly become nearly indistinguishable from the perspective of all local measurements. Related quantum information perspectives on chaos and thermalization have recently shed light on the physics of systems ranging from spin chains to black holes~ \cite{Larkin_1969_OTOC,Shenker2014,Shenker:2014cwa,Sachdev2015,kitaev2015,Maldacena2016,Aleiner2016,Hosur2016,Leviatan2017,Hallam:2019npn,White2017,Gu2017,Zhou2017,Luitz2017,Patel2017,Nahum2017a,Nahum2017a,VonKeyserlingk2017,Bohrdt2017a,Xu2018,Sahu2018,You2018,Nahum2018,Heyl2018,Dag_2019_OTOC}.


In this work, we consider spatially local systems and discuss a setup that can be used to sharply measure the way initially localized quantum information spreads in space as a function of time. Given a family of spatial subregions of fixed shape and variable size $R$, we diagnose information spreading by determining how $R$ must increase with time such that a subregion of size $R$ just contains some initially localized quantum information. For the class of models considered here, we find that information spreads ballistically, at an information speed $v_I$. This speed can depend on the shape of the regions used to define it, and we give a general criterion determining $v_I$ for any valid family of regions. Our focus is on strip-like regions where $R$ is taken to be the half-width of the strip. In this case, we find that the information speed is precisely related to two other characteristic speeds, the entanglement speed $v_E$, which measures how rapidly entanglement grows after a quantum quench~\cite{Liu:2013iza}, and the butterfly speed $v_B$, which measures how rapidly perturbations spread in space~\cite{Roberts_2015_vB,Roberts:2016wdl}.


The setup we consider is a quench experiment with two ingredients. First, given some spatially local chaotic system, we prepare an initial pure state $|\psi\rangle$ of some given energy density $\varepsilon$ above the ground state. This state may be out of local equilibrium, but chaos will lead to a late time state that is effectively thermal at some temperature $T(\varepsilon)$. At the initial moment, the entanglement of any region $A$ is taken to be a fixed fraction $f$ of its thermal value, $S(A) = f s |A| + \cdots$ where $0\leq f \leq 1$, $s$ is the thermal entropy density, and $\cdots$ denotes subleading terms. 

Second, we take an auxiliary qubit (or set of qubits) and maximally entangle it locally with the system. This extra system is called the reference; it undergoes no dynamics, except for this initial entangling operation. Now the system is allowed to evolve for a time $t$. The goal is to recover the entanglement with the reference using as small a system region as possible. 

Any subsystem whose mutual information with the reference is nearly maximal can recover the information. However, the size of the smallest such region is expected to grow with time as the local information spreads. Here we restrict to models with no chirality or preferred direction and consider the asymptotic limit of large regions and times. Using strip regions to define the information speed, we find
\begin{equation}
    v_I = \frac{v_E(f,\varepsilon)}{1-f}.
    \label{eq:main}
\end{equation}
Here $v_E(f,\varepsilon)$ is the entanglement speed, which measures the rate of entanglement growth for a strip region starting from an initial state of energy density $\varepsilon$ and entanglement fraction $f$. The energy dependence of various speeds will be mostly suppressed. We note that the prior literature on entanglement growth has focused primarily on the case $f=0$, so what is commonly called the entanglement speed is the single number $v_E(f=0)$ in our notation. Note also that the rate of entanglement growth can depend on the region shape. 

Eq.~\eqref{eq:main} is one of our main results; the evidence for it comes from three sources, a general quantum information argument, a microscopic spin chain calculation, and holographic calculations using AdS/CFT. As such, we conjecture that it holds generally for translation invariant chaotic quantum systems ranging from spin chains to black holes. As a function of $f$, we find that $v_I$ ranges from $v_E(f=0)$ at $f=0$ to $v_B$ at $f=1$. Hence, our setup unifies the entanglement speed $v_E(0)$ and butterfly speed $v_B$ by showing that they are particular limits of a general information speed. We also find that the various velocities depend only on energy density and other conserved quantities and, in the case of $v_E$, on the entanglement fraction $f$.

We emphasize that the above formula applies only to strip regions. This is because the quantum information argument leading Eq.~\eqref{eq:main} assumes that the time for entanglement entropy of a strip to saturate is set by the speed $v_{E}(f)$ as opposed to any other speed. This assumption is not true for arbitrary regions shapes, and we will discuss other subregion shapes at the end of the paper.


In the case of strips, we argue for a more general formula of the form
\begin{equation}
    v_I = \min \left( \frac{v_E(f,\varepsilon)}{1-f}, v_B(\varepsilon) \right),
    \label{eq:main_alt}
\end{equation}
but, when we can calculate reliably, we find
\begin{equation}
    \frac{v_E(f)}{(1-f)} \leq v_B
\end{equation}
and
\begin{equation}\label{eq:vE_limit}
    \lim_{f\rightarrow 1} \frac{v_E(f)}{1-f} = v_B.
\end{equation}
In particular, the statement that $v_E(f) \sim v_B (1-f) + \cdots$ near $f=1$ has appeared independently in recent membrane models of entanglement growth~\cite{Mezei2018,Jonay2018}.

From the perspective of our work, these statements would follow if the information speed were a non-decreasing function of the fraction $f$. This conjecture seems plausible since, as discussed below, more entanglement should only make information transmission easier. In App.~\ref{app:vE_bound} we give a precise but non-rigorous physical argument for the inequality $v_E(f) \leq (1-f) v_L$ where $v_L$ is an effective lightcone speed at a given energy density; under the assumption that $v_L=v_B$, which we also argue for, the above bound follows. We note that the coarse-grained entanglement growth model in Refs.~\cite{Mezei2018,Jonay2018} also implies such an inequality.

The key idea underlying Eq.~\eqref{eq:main} is a generalization of the Hayden-Preskill protocol \cite{Hayden:2007cs}, which enables us to track where the information is located as a function of time. That protocol has two key ingredients: a highly entangled state and a scrambled output. The appearance of the minimum over $v_E$ and $v_B$ in Eq.~\eqref{eq:main_alt} arises because one or the other of these ingredients serve as a bottleneck to information flow. The reason why the simpler Eq.~\eqref{eq:main} is obtained is that entanglement generation appears to always be the slower process in the cases we considered.

Interpreted directly, these results apply to the spreading and complexification of information. They do not immediately imply that the information can be read out locally. Indeed, this is generally thought to be impossible in chaotic systems, unlike in weakly interacting systems where information can propagate in localized wavepackets. However, with the right notion of signaling, we show that information does actually move at the speed $v_I$, although the decoding operation may be quite complex. We also discuss a local variant of the recent traversable wormhole setup~\cite{Gao:2016bin,Maldacena:2017axo} in the limit $f=1$ and show that, for this particular situation, there is a coherent local signal that moves at speed $v_I(f=1) = v_B$. The possible extension of these wormhole results to more generic models and away from $f=1$ is an interesting open question.

We now outline in somewhat more detail the main results of the paper. In section~\ref{sec:qi}, we give a general quantum information theoretic argument for Eq.~\eqref{eq:main} based on two streamlined assumptions about entanglement growth and operator growth. The main technical tool is a variant of the Hayden-Preskill protocol. The key result is that any region centered on the initial location of the quantum information can recover said information provided the region's entanglement entropy has not yet saturated. Hence, the information velocity is 
\begin{equation}\label{eq:ent_sat}
    v_I =\lim_{t\rightarrow \infty} \frac{R_{\text{sat}}(t)}{t},
\end{equation}
where $R_{\text{sat}}(t)$ is the size of the region whose entanglement has just saturated at time $t$. Assuming the entanglement velocity sets the saturation time for strips, then Eq.~\eqref{eq:main} follows immediately. Of course, in general, the entanglement of a region will not exactly saturate at any fixed time, but we expect that generalizing to the case of near saturation does not change the asymptotic scaling.

In section~\ref{sec:sc}, we calculate the mutual information between system subregions and the reference using a numerical Krylov technique applied to a spin chain with $22$ sites at energy densities corresponding to infinite temperature. The information speed is directly measured by extracting the level sets of the mutual information as a function of system size and time. The butterfly speed and the entanglement speed are also independently measured. Combining these, we find excellent agreement with Eq.~\eqref{eq:main} for small $f$ and qualitative agreement for larger $f$. We expect a significant finite-size effect on the measurement of $v_E$ at large $f$, so our results are consistent, given the system sizes accessible to us.

In section~\ref{sec:adscft}, we compute the information speed for strip-regions in any dimension as a function of temperature and charge density using planar charged Vaidya-AdS-Reissner-Nordstrom black holes. This is done by computing HRT surfaces~\cite{Hubeny_2007_Covariant_RT} to determine which boundary regions have entanglement wedges~\cite{Czech_2012_Entanglement_Wedge,Bousso_2012_Lightsheets_Subregions,Dong_2016_Entanglement_Wedge,Cotler_2017_Entanglement_Wedge} that contain an infalling particle that is entangled with the reference. The charge is a convenient additional handle, which allows us to tune the ratio of $v_E(0)$ to $v_B$. The entanglement fraction is incorporated by studying a thermofield double state at some pre-quench energy density, which is quenched to a higher energy density by a unitary perturbation. Our calculations are a significant generalization of recent calculations corresponding to the $f=1$ case in CFTs~\cite{Mezei_2017_Entanglement_Chaos}. With this setup, we show that both Eq.~\eqref{eq:main} and Eq.~\eqref{eq:ent_sat} are valid in AdS/CFT. We are also able to compute $v_E(f)$ and verify that Eq.~\eqref{eq:vE_limit} holds. While we focus on strips in any dimension, the results of this section could be easily generalized to more general shapes.

In section~\ref{sec:transmission}, we discuss the problem of information transmission and its relation to scrambling. Turning again to Hayden-Preskill, we show that information can, in some sense, be transmitted at the speed $v_I$. More precisely, if some information is initially encoded locally at position $x=0$, then after a time $t$ it will be spread over a region of size $v_I t$ but we only need a small bit of the region $|x| < v_I t$ plus the rest of the system to decode the information. We also generalize the recent traversable wormhole story in AdS/CFT to the spatially local case with $f=1$. In this context, we show that there is even a locally detectable signal moving at speed $v_I(f=1)=v_B$, at least at large $N$.

Finally, in section~\ref{sec:discussion} we summarize our results and discuss numerous extensions and open questions raised by our work.


\section{Quantum information argument}\label{sec:qi}

This section develops a general quantum information argument for the relation in Eq.~\eqref{eq:main} for strip regions for a broad class of chaotic systems. The argument focuses on the asymptotic limit of large sizes and times, although the discussion can be refined to include finite-size effects. For simplicity, we consider a one-dimensional parity-symmetric chaotic system, although generalizations to higher dimensions are possible.\footnote{One subtlety in higher dimensions (or without the parity assumption) is that the information speed can be anisotropic. Relatedly, one also needs to consider shape dependence.}

\subsection{Assumptions and tools}

The first main assumption concerns entanglement growth. Let $|\psi\rangle$ be a translation-invariant pure state of energy density $\varepsilon$ above the ground state. The general expectation in a chaotic system is that any two such states behave identically as far as local probes are concerned, except at early time before local equilibrium has been reached. For our purposes, it suffices to assume that such states feature volume law entanglement and linear-in-time entanglement growth. More precisely, the entropy of a simply-connected region $A$ is assumed to be
\begin{equation}
    S[A,\psi(t)] = \min\left\{ f s(\varepsilon) |A| + v_E(\varepsilon,f) s(\varepsilon) |\partial A| t, s(\varepsilon) |A|\right\}
\end{equation}
where $s(\varepsilon)$ is the thermal entropy density at energy density $\varepsilon$ and $v_E(\varepsilon,f)$ is the above entanglement speed.

The form of $S[A,\psi(t)]$ assumed above means in particular that saturation occurs immediately after the linear growth. Thus, the saturation is a first-order transition since the time derivative of the entanglement entropy jumps discontinuously from a nonzero value to zero. The assumption above excludes the case where $A$ is a spherical subregion, for example. For concreteness, we continue to focus on strips but will revisit the more complicated case of spheres in subsection \ref{subsec:sphere} later on.

Hence, the states of interest exhibit possibly sub-thermal (corresponding to $f < 1$)  volume law entanglement at early time and a constant entanglement growth rate until saturation. More realistically, one expects subleading in system size corrections to the entropy and more complicated time-dependence. However, the volume law component of the entanglement and the period of constant entanglement growth are expected to dominate the physics at large size and long time.

The second main assumption concerns chaos and operator growth. To set the stage, the local physics is assumed to be describable in terms of a model of ``thermal cells''. There is a length scale $\xi(\varepsilon)$ such that local observables are approximately uncorrelated beyond range $\xi(\varepsilon)$. Chunks of linear size $\xi$ are described by a Hilbert space of size $e^{s\xi}$ (in one-dimension), and the thermal dynamics takes place in the tensor product space built from these local cells.

Consider a local perturbation created by an operator $W$ restricted to a single thermal cell. The growth of the perturbation can be measured using the size of the Heisenberg operator $W(t)$. We assume that this operator acting on a state of a given energy density is approximately supported on an interval of length $2 v_B(\varepsilon) t$. This means that on any state of the same energy density, we may replace
\begin{equation}
    W(t) |\psi\rangle \rightarrow \tilde{W}(t) | \psi\rangle
\end{equation}
where $\tilde{W}(t)$ is supported on an interval of length $2 v_B t$ or, equivalently, on $\frac{2 v_B t}{\xi}$ thermal cells. The spreading is assumed to be uniform in a precise sense discussed below.

We will assume $v_E \leq v_B$, as this is true in all known cases. However, this inequality is not rigorously proven; for Lorentz-invariant field theories, the weaker statement that $v_E \leq c$, with $c$ the speed of light, is proven~\cite{Hartman_2015_Bound_vE,Casini_2016_Bound_vE}. We assume that $v_B$ functions like the effective maximum speed limit at the given energy density~\cite{Roberts:2016wdl}.

\begin{figure}
    \centering
    \includegraphics[width=.9\textwidth]{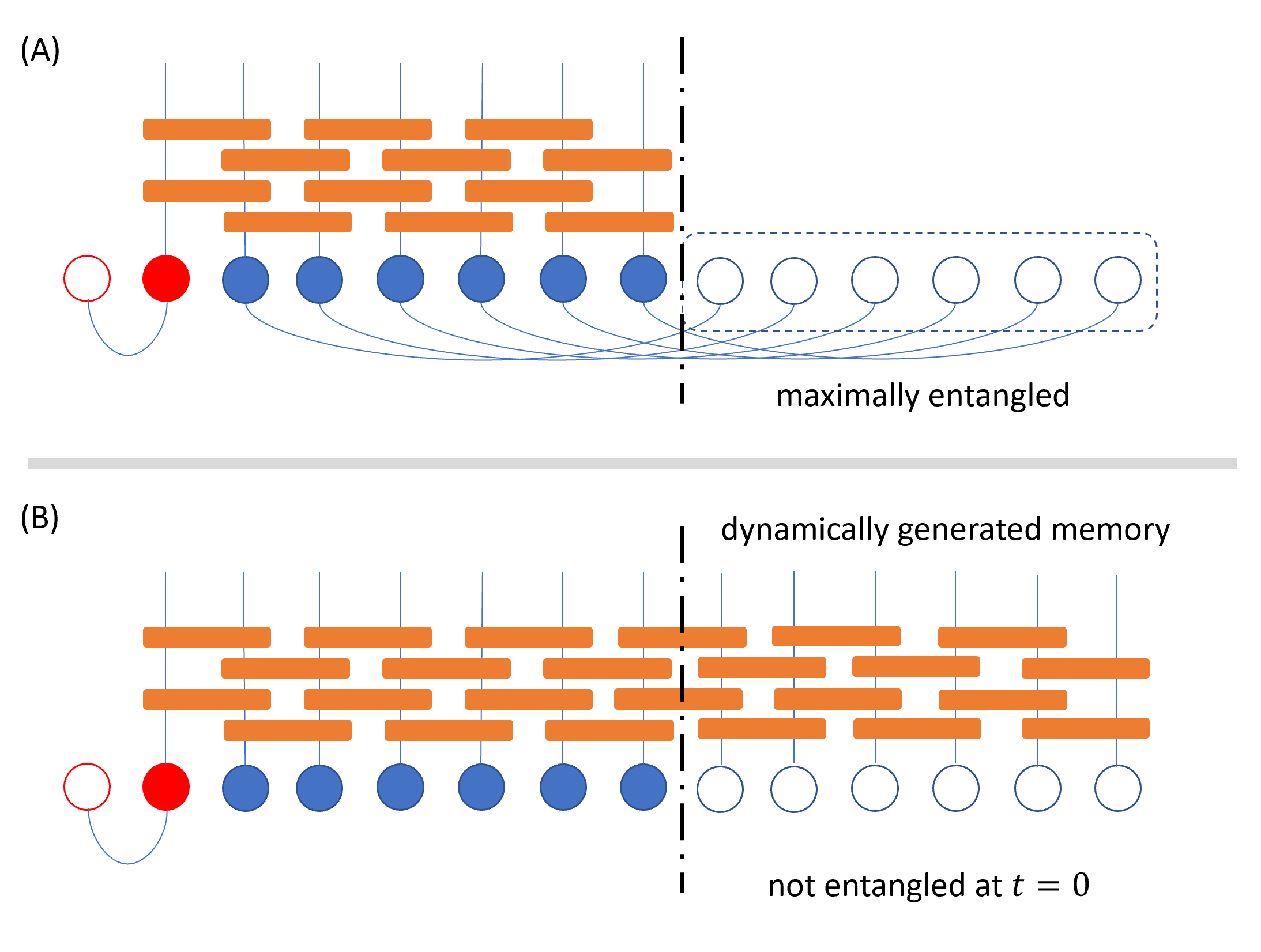}
    \caption{\textbf{Panel A (top):} Conventional Hayden-Preskill protocol for a spatially local scrambler. The right side of the dot-dashed line is the memory, which is maximally entangled with the system at time zero and frozen thereafter. The left side of the dot-dashed line contains the system which undergoes local scrambling unitary dynamics. Once the information correlated with the reference (red circle) spreads over the whole system, access to the memory plus a few bits of the system suffice to recover the entanglement with the reference.\\ \textbf{Panel B (bottom):} For $f=0$ (or more generally $f<1$), there is no (or submaximal) entanglement with the memory. Instead, the local scrambling dynamics will also dynamically generate the memory in addition to spreading the information. The bottleneck step turns out to the generation of the memory, hence relating the success of this generalized Hayden-Preskill protocol to the generation of maximal entanglement. }
    \label{fig:hp_explain}
\end{figure}

Given these assumptions, the main technical tool below is a generalization of the Hayden-Preskill protocol. This protocol is illustrated in Figure~\ref{fig:hp_explain}; the top panel illustrates the conventional Hayden-Preskill protocol, while the bottom panel shows our generalized Hayden-Preskill protocol. Conventional Hayden-Preskill considers three systems, a reference, a system, and a memory. After an initial entangling operation, the system is the only party that has non-trivial dynamics. One then finds that once the system is scrambled, access to the memory plus a few bits of the system suffice to recover the information.

The simplest heuristic for this phenomenon comes from the no-cloning theorem. Suppose a reference is maximally entangled with a system in some complex quantum state. Then we should be able to recover the entanglement with the reference from any part of the system that contains a little more than half the degrees of freedom. If we could get away with fewer than half, then we could recover the entanglement in two different places simultaneously, violating no cloning. The whole memory plus a little bit of the scrambled system is more than half the degrees of freedom of system plus memory, so we expect recovery of the reference entanglement to be possible.

Relative to conventional Hayden-Preskill, the key difference in our case is that, except when $f=1$, some or all of the ``memory'' which is entangled with the ``system'' must be dynamically generated. Indeed, the ``system'' and ``memory'' are all part of one large evolving system. Hence, even when maximal entanglement has been generated, it will not be of simple Bell pair type as in conventional Hayden-Preskill. However, this will not affect the criterion for decoding since the distillation to Bell pairs can be done entirely within the memory part (the complexification of the entanglement will likely make the decoding more difficult).

\subsection{Argument}

Now for the main argument. Consider an initial state $\psi_0$ which has entanglement fraction $f$. All quantities considered below can depend on energy density, but the explicit argument is suppressed for notational simplicity. At time zero, the system is locally entangled with a reference system $\text{REF}$ consisting of $|\text{REF}|$ qubits. This entanglement is produced by generating $2^{|\text{REF}|}$ orthogonal states using local operators restricted to a few thermal cells near the origin of space, $x=0$. Although $\text{REF}$ can be more than a single qubit, the entangling operations should not significantly change the energy density. However, some local energy may be added to increase the number of active local degrees of freedom.

After the initial entangling operation, the reference is set aside, and the system is allowed to evolve for a time $t$. The question, as outlined in the introduction, is what is radius $R_I(t)$ of the smallest sized interval (of total width $2R_I$) centered at the origin which can just recover the entanglement with the reference. Some region $A$ can recover the entanglement if the mutual information is nearly maximal,
\begin{equation}
    I(A:\text{REF}) = 2|\text{REF}| \rightarrow \text{recovery possible}.
\end{equation}
Throughout the argument, $A$ denotes an interval of radius $R$ centered at the origin.

Consider first the case of $f=1$, so that the initial state is maximally thermally entangled and in local thermal equilibrium on all scales. By assumption, $v_B$ is the maximum speed limit; hence no information can travel faster than $v_B$, so $R_I \leq v_B t$. This statement should be understood as being valid on the scale of a few thermal cells.

To show that actually $R_I = v_B t$, we argue that $R_I \geq v_B t$. This is done by asking under what conditions the complement $A^c$ of a region $A$ can recover the entanglement. Because the overall state is pure, the mutual information obeys
\begin{equation}
    I(A:\text{REF}) + I(A^c:\text{REF}) = 2 |\text{REF}|,
\end{equation}
so if $I(A:\text{REF})=0$ then necessarily $A^c$ can recover the entanglement with $REF$ and vice versa. Because $v_B$ sets the growth speed of local operators and because the initial state is highly entangled, the Hayden-Preskill protocol implies that the information can be recovered in the complement if $R < v_B t$. 

Here is the precise argument. Let $\Delta A$ denote a chunk of $A$ consisting of a few thermal cells. If the radius of $A$ is less than or equal to $v_B t$, then this chunk contains some of the output of the chaotic dynamics that has scrambled the initial entanglement. As such, if this piece is sent to the complement of $A$, then because $A^c$ and $A$ are maximally entangled, the Hayden-Preskill protocol implies that the entanglement is now recoverable in the augmented complement, $A^c + \Delta A$. Hence, $R_I \geq v_B t$. Combined with the upper bound $R_I \leq v_B t$, we learn that $R_I = v_B t$ when $f=1$. 

The equality $v_I = v_B$ will continue to hold for $f<1$ if $v_B \leq v_E(f)/(1-f)$ since the relevant regions will become maximally entangled fast enough for operator growth to remain the bottleneck process. However, we should emphasize that, apart from some unreliable large $f$ spin chain numerics, we do not have an example where $v_B < v_E(f)/(1-f)$ for $f<1$. It may be impossible.

More generally, when $f \neq 1$, it becomes necessary to track both the location of the chaotic output and the degree of entanglement. This is because success in Hayden-Preskill requires both access to the output and maximal entanglement. It is now possible to recover the entanglement in $A$ even when $R < v_B t$ (assuming $v_E(f) < v_B (1-f)$) because even though the complement holds some of the chaotic output, it does not yet have enough entanglement with $A$ to use the Hayden-Preskill protocol.

In fact, assuming $v_B > v_E(f)/(1-f)$, the minimum size region which can recover the entanglement at a given time $t$ is the one whose entanglement is just about to saturate. To show this, consider a family of regions of different sizes centered on the initial location of the information.

For this class of regions, we need to calculate the effective size of the system plus memory in Hayden-Preskill. At a given moment in time, the overall system has dynamically generated an effective ``system'' and ``memory'' for the purposes of Hayden-Preskill. The effective size of the ``system'' plus ``memory'' is equal to twice the size of the largest region whose entropy has just saturated. The region that has just saturated is the ``system'' and the part of the complement it is entangled with is the ``memory''. Then any region whose entropy has not yet saturated has access to both the whole ``system'' and part of the ``memory'' and hence can recover the information. By contrast, if a region has saturated its entropy, it contains only part of the ``system'' and none of the ``memory,'' and hence does not have access to the information (because its complement does, by Hayden-Preskill).

The saturation time for a region of size $R$ is
\begin{equation}
    t_{\text{sat}} = R(1-f)/v_E(f),
\end{equation}
so the information velocity is
\begin{equation}
    v_I = \frac{R}{t_{\text{sat}}} = \frac{v_E(f)}{1-f}.
\end{equation}
Note that while the discussion here focused on pure states, it can be straightforwardly generalized to mixed states which are purified by some other system, distinct from the reference. In this case, recovering information in the complement typically requires access to part of the purifying system as well.

\section{Spin chain calculations}
\label{sec:sc}

To be concrete, we study the above information protocol in a context of spin chains with local interaction. Consider the following initial state,
\bea
\ket{\Psi}= \frac{1}{\sqrt{2}}\left(\ket{\uparrow}_\text{REF} \ket{\psi_\uparrow} + \ket{\downarrow}_\text{REF} \ket{\psi_\downarrow}\right)
\label{eq:spin_init}
\eea
where the two states in the system $\psi_{\uparrow/\downarrow}$ are orthonormal so that the reference spin $\text{REF}$ is maximally entangled with the system. Initially, we set the two states as,
\bea
\ket{\psi_{\uparrow,\downarrow}}= P_{\uparrow,\downarrow} \ket{\Psi} \eea
with proper normalization, 
where $P$ projects the local state at site 0 to spin up or spin down.
As a result, the reference is entangled with the system through the first spin. The dynamics of the system is governed by a local Hamiltonian $H$, which does not affect the reference. As time increases, it is expected that more and more spins start to participate in the entanglement between the reference spin and the system and that the local information spreads out. The entanglement expansion is tracked by the mutual information,
\bea
I(\text{REF},A)&=S_{\text{REF}}+S_A-S_{\text{REF}\cup A}=S_{\text{REF}}+S_A-S_{\bar{A}}.
\eea 
where $\bar{A}$ labels the compliment of A. The reduced density matrix of region $A$ is given by
\bea
\rho _A = \frac{1}{2} \tr_{\bar{A}} \left ( \ket{\psi_\uparrow} \bra{\psi_\uparrow} + \ket{\psi_\downarrow} \bra{\psi_\downarrow}\right ),
\eea
similar expression for $\rho_{\bar{A}}$, from which $S_A$ and $S_{\bar{A}}$ can be calculated. 

To study how the information velocity depends on the entanglement density of the initial state,  we prepare the initial state for the $L$ sites spin chain as,
\bea
\ket{\Psi} =\prod_{r=1} ^{L/2}  a_{\sigma \sigma'} \ket{\sigma_r} \ket{\sigma'_{r+L/2}}.
\label{eq:f_state}
\eea
where the coefficient $a$ is chosen so that the entanglement between the spin on site $r$ and site $r+L/2$ is $f\log 2$.  
The state $\ket{\Psi}$ is a generalization of $L/2$ Bell pair, in which the entanglement between two spins in a pair is not maximal but $f \log 2$. As a result, $\ket{\Psi}$ has the required entanglement density $f$.

After discussing the general set-ups, we study a specific example where the Hamiltonian is given by the following mixed-field quantum Ising model in one dimension,
\bea
H&=-\left ( J \sum\limits_{r=1}^{L-1} Z_r Z_{r+1} + h_z\sum\limits_{r=1}^L Z_r + h_x\sum\limits_{r=1}^L X_r   \right ) \\
\label{eq:ham}
\eea
where $Z_r$ and $X_r$ are local Pauli operators. We set the parameters $J=1$, $h_x=1.05$, and $h_z=0.5$. This model is generally non-integrable and has been widely used to study quantum chaos and thermalization. In this problem, the coefficient $a$ in Eq. \ref{eq:f_state} is set to
\bea
a(f) = \left(\frac{1}{2}I +\xi Y \right ) ^{1/2}
\eea
where $\xi$ satisfies $-(1/2+\xi)\log (1/2+ \xi)-(1/2-\xi)\log (1/2- \xi) = f\log(2) $. With this set-up, the local density matrix is orthogonal to the energy density operator. In other words, initial states with different f all have total energy 0. They are all in the middle of spectrum and their entanglement entropy after long-time unitary evolution is only bounded by the local Hilbert space dimension. 

\begin{figure}
\includegraphics[height=0.45\columnwidth, width=0.45\columnwidth]
{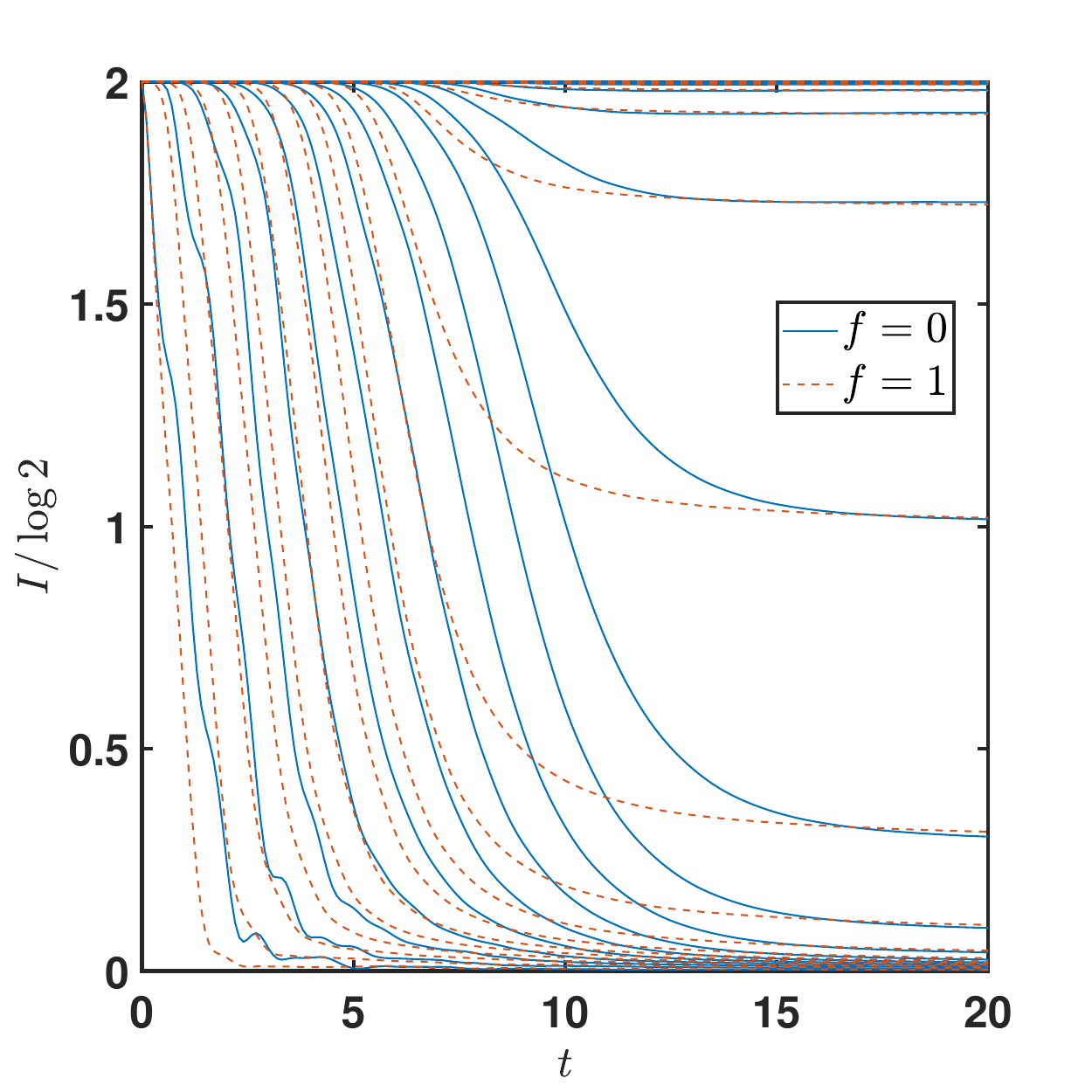}
\includegraphics[height=0.45\columnwidth, width=0.45\columnwidth]
{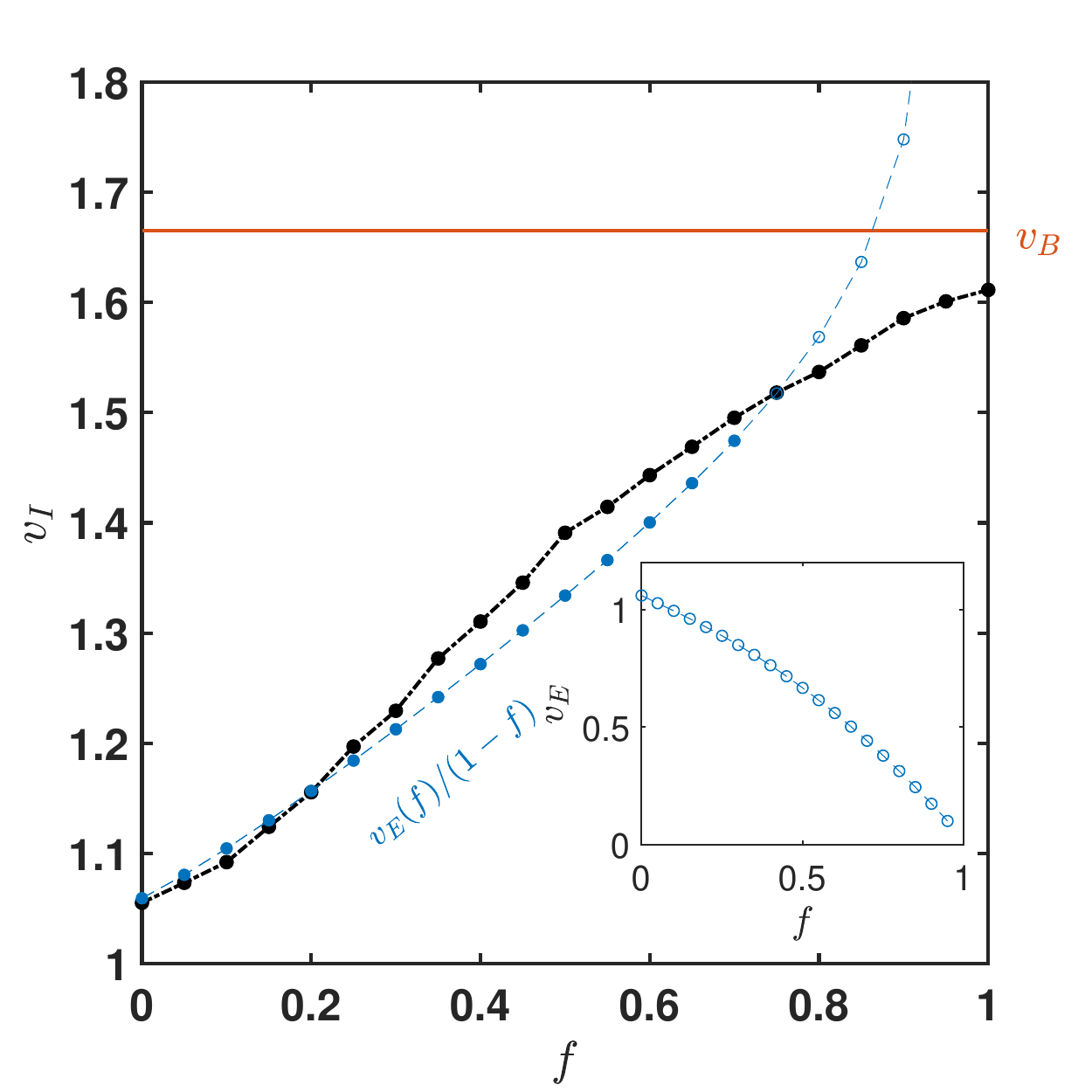}
\caption{(a) Comparing the information expansion between a product state and a maximally entangled state. Evidently, the information expansion in the maximally entangled case is faster. (b) The information speeds increases with the entanglement density $f$, up to approximately the  butterfly speed as $f$ goes to 1 and is tracked by $v_E(f)/(1-f)$. The entanglement speed as a function $f$ is shown in the inset.}
\label{fig:mut}
\end{figure}

We apply the information protocol for systems with 22 spins. The unitary time evolution is implemented by the standard Krylov subspace method. In practice, we put the initial entangled spin on the boundary to allow for more space for information expansion. The mutual information between the reference and subsystem with increasing length is shown in Fig. \ref{fig:mut}(a) for $f=0$, i.e., a product state and $f=1$, a maximally entangled state. The comparison between the two cases clearly indicates that information transport is slower for small $f$, consistent with the general argument. We extract the information velocity by a linear fit on the wavefront of the mutual information and plot it as a function of $f$ in Fig. \ref{fig:mut}(b). As $f$ increases, the information velocity increases up to approximately the butterfly speed from the out-of-time ordered correlator. Furthermore, the $f$ dependence of $v_I$ is captured by the general formula Eq. \ref{eq:main}. When $f$ is large, it is difficult to extract $v_E$ accurately because the finite size only allows limited room for the entanglement growth before saturation, and any inaccuracy is amplified by $(1-f)^{-1}$ when comparing Eq. \ref{eq:main} to $v_I$, leading to the deviation. 

In condensed matter systems, the various speeds are usually difficult to calculate analytically. One exception is the random circuit model in the large spin limit, where it is shown that $v_B=1$~\cite{Nahum2017a,VonKeyserlingk2017}, only restricted by geometry structure of the unitary operators, and $v_E(f) = (1-f)^{-1}$~\cite{Nahum2017,Jonay2018,Zhou2018}, obtained from the subadditivity of the entanglement entropy. Applying the formula \ref{eq:main}, we obtain that $v_I=v_B=1$, regardless of the entanglement fraction of the initial state. This meets the expectation that the information should propagate as fast as possible in this case, providing further support for our theory. Exploring the information speed at finite q would be an interesting future research direction.  

\section{Holographic calculations: $v_{I}$ from HRT surfaces}
\label{sec:adscft}

We now turn to calculations using AdS/CFT to provide further evidence in support of formula (\ref{eq:main}).  

\subsection{Information velocity from HRT surfaces}
The scrambling of the entanglement between the system and the reference $\text{REF}$ is captured holographically by letting a particle entangled with $\text{REF}$ fall into the bulk from the boundary. The particle is contained inside the entanglement wedge of some boundary region if and only if the entanglement between the system and $\text{REF}$ is recoverable inside that boundary region. This follows from a direct computation of the mutual information using the quantum corrected RT formula~\cite{Faulkner_2013_Entanglement_RT_Quantum}. As the particle falls further and further into the bulk, the smallest possible entanglement wedge that contains the particle grows in size, reflecting the boundary statement that the system-reference entanglement is scrambled over a larger and larger boundary region. We will study two examples of bulk spacetimes:\\

(1) a planar black hole formed from the collapse of a thin, electrically charged shell, i.e., the single-sided AdS-Reissner-Nordstrom-Vaidya solution, and\\

(2) a charged, collapsing thin shell on top of the eternal AdS-RN black hole, i.e., the 2-sided AdS-RN-Vaidya solution.\\

In each case, we let a particle fall into the bulk, and we find numerically the smallest entanglement wedge containing that particle at different times on the boundary. For simplicity, we will specialize to the case where the boundary region is a strip with some extent $[-R, R]$ along the $x^{1}$-direction, and with the other $x^{i}$ directions ranging from $-\infty$ to $\infty$. The rate of expansion of that smallest entanglement wedge gives us $v_{I}$, which is the left-hand side of equation (\ref{eq:main}):
\begin{equation}
    v_{I} = \mathrm{min}{\left( \frac{v_{E}(f)}{1-f}, v_{B} \right)}
\end{equation}
As for the right-hand side, $v_B$ is known independently, and in the case of the single-sided Vaidya solution, where $f=0$, $v_{E}$ is also known analytically (see the next paragraph). In the case of the two-sided Vaidya solution, we employ analytic expressions derived in appendix \ref{App:ve} for $v_E(f)$, as well as perform an independent numerical study of the rate of growth of the entanglement entropy of boundary strips with a fixed (but large) width as a consistency check of $v_{E}(f)$. Using both methods, we compare the left-hand side with the right-hand side for $f$ ranging from $0$ to $1$.

We have chosen to study strips since this is the simplest shape consistent with the assumption of a first-order phase transition stated in section \ref{sec:qi}. For $f=0$, this is known from the previous studies \cite{Liu:2013iza, Liu:2013qca}. For $f>0$, that the transition is first-order is a conclusion of our work. The case of spherical subregions will be treated in section \ref{subsec:sphere}.

The reason why we include electric charge is that the near-extremality limit allows for a parametric separation between the two velocities $v_{E}$ and $v_{B}$, which allows for cleanly verifying formula (\ref{eq:main}). To see this parametric separation, we first record the formulae for $v_{B}$ and $v_{E}$ for the AdS-RN black hole in $(d+1)$ dimensions. The butterfly speed is given by \cite{Blake:2016jnn,Roberts:2016wdl}:
\begin{equation}\label{vBAdSRN}
    v_{B} = \sqrt{\frac{2\pi z_{+} T}{d-1}}
\end{equation}
where $T$ is the temperature and $d$ is the boundary spacetime dimension (and we have set $L=1$). The entanglement speed is given by \cite{Liu:2013iza, Liu:2013qca}:
\begin{equation}\label{vEAdSRN}
    v_{E} = \sqrt{\frac{1}{\eta-1}} \left( \left( 1 - \frac{u}{\eta} \right)^{\eta} - (1-u) \right)^{1/2}
\end{equation}
with $\eta \equiv \frac{2(d-1)}{d}$ and $u \equiv \frac{4\pi z_{+} T}{d}$. We will focus on the near-extremality limit, since in this limit the two velocities above are parametrically separated. Indeed, $v_{B}$ scales as $\sqrt{T}$ near extremality, while $v_{E}$ scales as $T$. Thus, $v_{E}$ is much smaller than $v_{B}$ near extremality.\\

\subsection{Single-sided AdS-RN-Vaidya}\label{subsec:1sidedVaidya}
\begin{figure}
  \centering
    \includegraphics[width=0.4\textwidth]{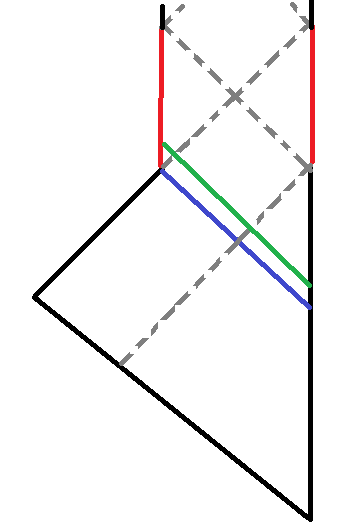}
      \caption{Penrose diagram of the planar, 1-sided AdS-RN-Vaidya spacetime. The infalling shell is in blue. The infalling particle is in green. The horizons are in dashed gray, and the timelike singularities in red.}
      \label{PenroseDiagram}
\end{figure}

Thanks to the parametric separation between $v_{E}$ and $v_{B}$ mentioned above, we expect that, for the single-sided AdS-RN-Vaidya solution:
\begin{equation}
    v_{I} = v_{E}
\end{equation}
and this is what we would like to check holographically. The metric in this case reads:
\begin{equation}\label{AdSRNVaidyametric}
    ds^{2} = \frac{1}{z^{2}} [-f(z,v)dv^{2} - 2dzdv + d\vec{x}^{2}]
\end{equation}
with
\begin{equation}
f(z,v) = \bigg \{ 
\begin{array}{ll}
    1 & v<0 \\
    1 - mz^{d} + q^{2}z^{2d-2}  & v>0 
\end{array} 
\end{equation}
We draw the Penrose diagram of this spacetime in figure \ref{PenroseDiagram}, together with the null trajectory of an infalling particle, which falls into the bulk at some time $t_{0}$ on the boundary. We take $t_{0} > 0$ so that the particle falls in after the shock does.\\

In this spacetime, there exists two qualitatively different types of extremal surfaces:\\
(1) the ones that pierce the infalling shell and poke into the empty AdS region, and\\
(2) the ones that stay entirely outside the infalling shell.\\
We will refer to the first type as O-type extremal surfaces (O for out-of-equilibrium), and the second type as E-type extremal surfaces (E for equilibrium). It might be that both an E-type surface and an O-type surface exist, which are anchored at the same boundary strip at the same boundary time. In that case, the HRT surface is the extremal surface with the smallest area. The terminology E- and O-type is motivated by the results of \cite{Liu:2013iza, Liu:2013qca}: this work found that - for a boundary strip of a fixed width - the HRT surface is of O-type at intermediate times (before thermalization), and the entanglement entropy grows linearly with time. At sufficiently late time, the HRT surface is of E-type, and the entanglement entropy has saturated to its final value.\\

\begin{figure}
\centering
\includegraphics[width=3in]{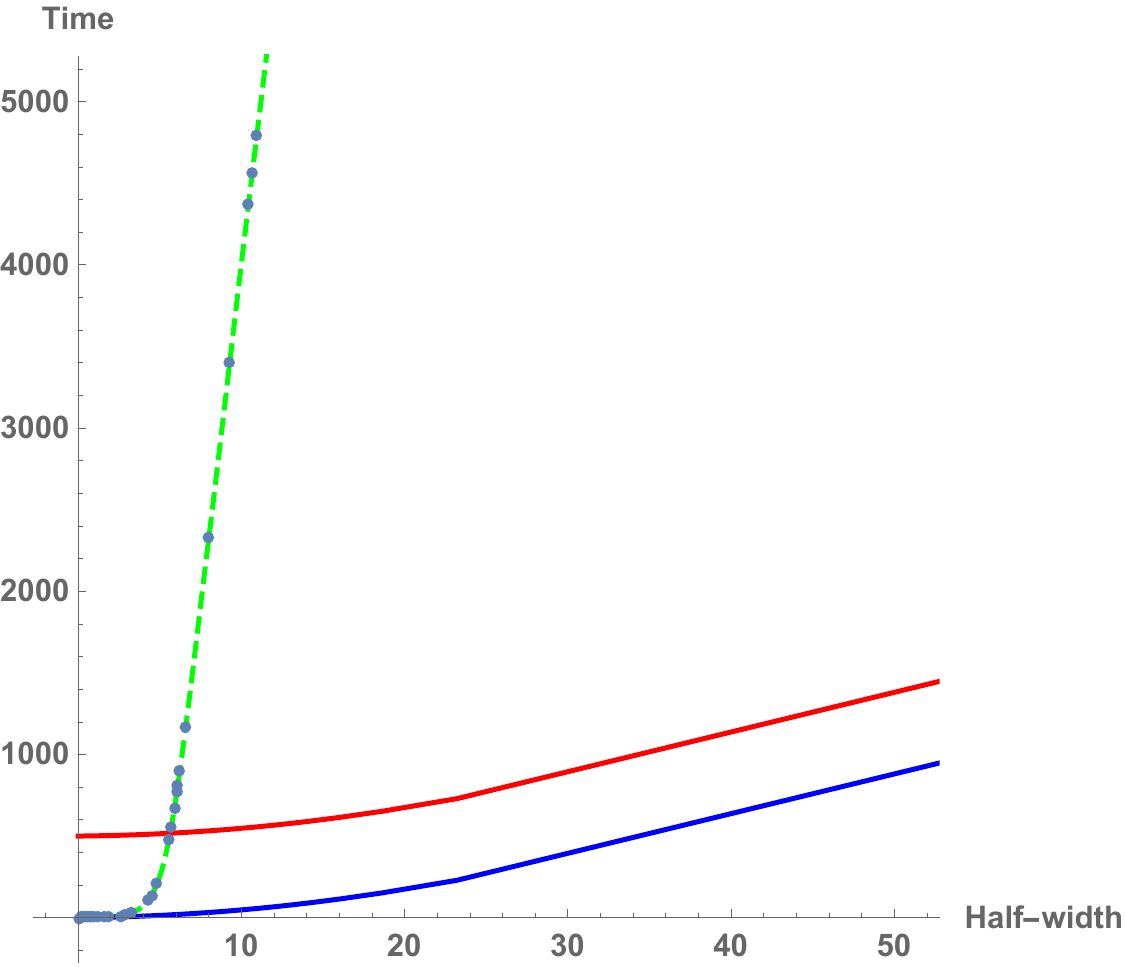}
 \caption{Phase diagram with the parameters $z_{+}=1$ and $z_{-}=1.001$. The blue dots lying on the green curve are numerically computed, and the green curve is the interpolation of the blue dots.}
 \label{1sidedPhaseDiagram}
\end{figure}

We relegate to \ref{App:ve} the technical details of how to find extremal surfaces. Appendix \ref{App:ve} works out the extremal surfaces in the more general case of the 2-sided Vaidya solution, but the single-sided case can be recovered as a special case (where $f=0$). We present the final numerical results in Figure \ref{1sidedPhaseDiagram} below in the form of a ``phase diagram''. The horizontal axis of the phase diagram is the half-width $R$ on the boundary, and the vertical axis is the boundary time $t_{b}$. Thus, points on the phase diagram scan over all possible boundary strips.\\

We numerically plot three curves (red, green, and blue), which contain information about whether the E- and O-type extremal surfaces exist for a given $R$ and $t_{b}$, and whether the HRT surface is of E- or O-type. The meaning of each of the three curves is as follows:
\begin{itemize}
    \item The red curve tracks the boundary strips whose E-type surface barely contains the infalling particle. The intersection of this curve with the vertical axis gives the boundary time at which the particle falls into the bulk. Varying this time of in-fall corresponds to shifting the red curve in the vertical direction. The reciprocal of the slope of the red curve at large $R$, large $t_{b}$ is the butterfly speed $v_{B}$.  
    \item The blue curve follows the largest possible half-width $R$, for a given $t_{b}$, such that an E-type surface exists. In other words, points on the blue curve are E-type surfaces whose tips lie on the infalling shell. If a point ($R$,$t_{b}$) on the phase diagram lies to the right of the blue curve, then there is no E-type surface corresponding to those values of $R$ and $t_{b}$.\\
    To see this, consider the eternal black hole geometry obtained by maximally extending the geometry to the future of the shell. Then the E-type surface in that geometry, which is anchored at ($R$,$t_{b}$), intersects the location of the trajectory of the infalling shell in the Vaidya geometry. Similarly, if a point ($R$, $t_{b}$) on the phase diagram lies to the left of the blue curve, then there exists an E-type anchored at that value of $R$ and of $t_{b}$.
    \item The dashed green curve represents boundary strips ($R$,$t_{b}$) such that the E- and O-type extremal surfaces both exist and have equal area. If a point ($R$,$t_{b}$) on the phase diagram lies to the right of the green curve (but to the left of the blue curve), then both the E- and O-type surfaces exist, and the HRT surface is of O-type. If a point ($R$,$t_{b}$) on the phase diagram lies to the left of the green curve, then both the E- and O-type surfaces exist, and the HRT surface is of E-type. 
\end{itemize}
From the phase diagram, we can infer the smallest entanglement wedge containing the infalling particle. That wedge is given by either the red or the green curve, whichever is closer to the vertical axis. When the particle has barely fallen into the bulk, the red curve lies closer to the vertical axis than the green one, and the smallest entanglement wedge containing the particle is of E-type. At later times, however, the green curve lies closer to the vertical axis than the red one, and the smallest entanglement wedge containing the particle is of O-type. Since we are only interested in the late-$t_{b}$, large-$R$ limit, we conclude that $v_{I}$ is the reciprocal of the slope of the green curve.\\
Numerically, $v_{E} \approx 0.00122291$ for our choice of $z_{+}$ and $z_{-}$, whereas the reciprocal of the slope of the green curve is found to be around $0.00121461$, agreeing with $v_{E}$ to within better than $1 \%$. Finally, we also note that the fact that the reciprocal slope of the green curve is $v_{E}$ can be derived analytically. We present that derivation in appendix (\ref{App:ve}) in the more general framework of the 2-sided AdS-RN-Vaidya solution. 

\subsection{2-sided AdS-RN-Vaidya}
Next, we study the case of 2-sided AdS-RN-Vaidya. The main advantage of this case, compared to the single-sided case, is that we can vary the entanglement fraction $f$ away from $f=0$. The metric is given by:
\begin{equation}
    ds^{2} = \frac{1}{z^{2}} [-h(z,v)dv^{2} - 2dzdv + d\vec{x}^{2}]
\end{equation}
with
\begin{equation}\begin{split}
h(z,v)&=h_i(z)+\Theta(v)\left(h_f(z)-h_i(z)\right)\\
&h_i(z)= 1 - m_{i}z^{d} + q_{i}^{2}z^{2d-2}\\
&h_f(z)= 1 - m_{f}z^{d} + q_{f}^{2}z^{2d-2}
\end{split}\end{equation}
and $\Theta(v)$ is a Heaviside step function. Here $m_{i}$, $q_{i}$ are the initial mass and charge, and $m_{f}$, $q_{f}$ are the final mass and charge. We will also denote by $z_{i,+}$ and $z_{i,-}$ the two roots of $h(z,v)$ before the shell, and $z_{f,+}$ and $z_{f,-}$ the two roots of $h(z,v)$ after the shell.\\
The equilibrium entropy density after thermalization is holographically given by:
\begin{equation}
    s_{eq} = \frac{1}{4G_{N}} \left( \frac{L}{z_{f,+}} \right)^{d-1}
\end{equation}
and the entanglement fraction $f$ is:
\begin{equation}\label{holographicf}
    f = \left( \frac{z_{f,+}}{z_{i,+}} \right)^{d-1}
\end{equation}
We keep fixed the geometry to the future of the shell in order to keep $v_{B}$ fixed (since $v_{B}$ only depends on the geometry to the future of the shell for a particle falling into the bulk after the shell does), and vary the geometry to the past of the shell in order to vary $f$.  Taking advantage of the fact that $v_B$ and $v_E(f=0)$ are parametrically separated near extremality, we choose the future geometry to be fixed $z_{f,+} = 1$ and $z_{f,-} = 1.01$.  For these values, equations (\ref{vBAdSRN}) and (\ref{vEAdSRN}) give $v_B\approx0.121$ and $v_E(f=0)\approx0.012$.\\
For each value of $f$, on the one hand, we produced a ``phase diagram'' similar to the one for the single-sided AdS-RN-Vaidya solution shown previously. From this phase diagram, we extracted $v_{I}$ from the slope of the green curve.
On the other hand, we compute the entanglement velocity through two methods.  First, we employ the results of appendix \ref{App:ve} that
\begin{equation}\label{eq:veAnalytic}
    v_E=\frac{z^{d-1}_{f,+}}{z^{d-1}_{i,+}}\sqrt{-h_f(z_m)\left(\frac{z^{2(d-1)}_{i,+}}{z_{m}^{2(d-1)}}-1\right)}
\end{equation}
with $z_m$ being the location of the local minimum of the function $h_f(z)\left(\frac{z_{i,+}^{2(d-1)}}{z^{2(d-1)}}-1\right)$ between $z_{f,+}$ and $z_{i,+}$.
As a consistency check, we also compute $v_E$ numerically through the following method. We first compute the entanglement entropy as a function of time for a boundary strip with a fixed but large half-width $R$. From this, we extract the value of $v_{E}$ corresponding to that value of $f$ from the rate of growth of the entanglement entropy at the moment of saturation. 

\begin{figure}
\includegraphics[height=0.45\columnwidth, width=0.45\columnwidth]
{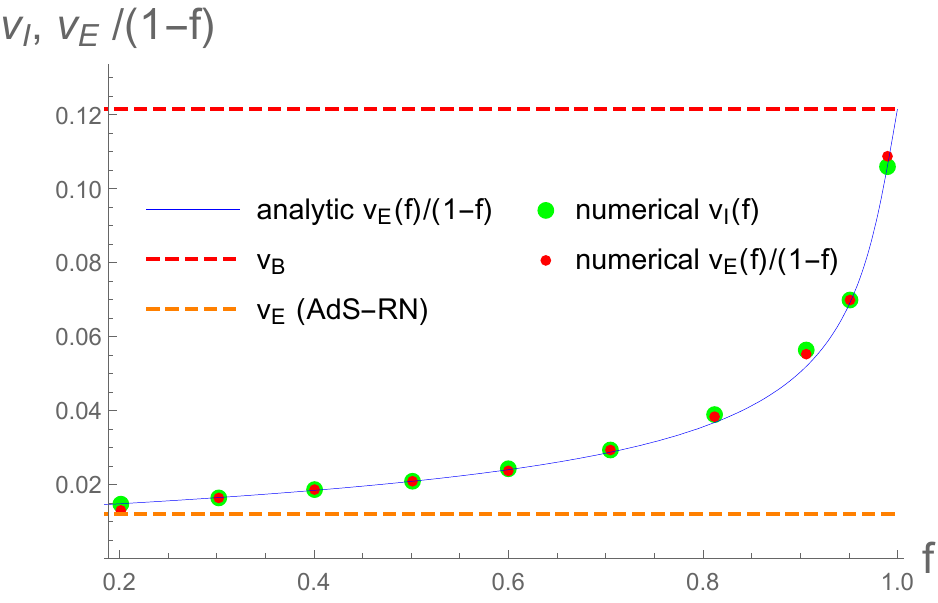}
\includegraphics[height=0.45\columnwidth, width=0.45\columnwidth]
{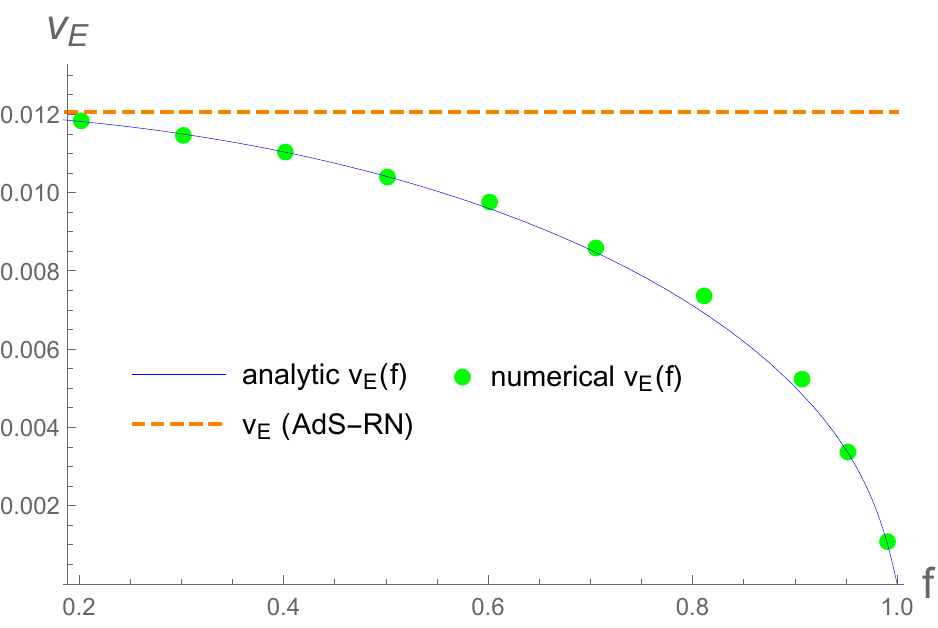}
\caption{Left panel: Red and green dots are $v_{I}$ and $v_{E}(f)/(1-f)$, respectively, computed numerically for a range of values of the entanglement fraction $f$. The blue line is $v_{E}(f)/(1-f)$ as obtained through equation (\ref{eq:veAnalytic}). Right panel: Green dots show numerically obtained values of $v_{E}$, while the blue line is $v_{E}(f)$ from equation (\ref{eq:veAnalytic}).}
\label{fig:2sidedVaidyaplot}
\end{figure}

With these results in place, we then compare the two sides of equation (\ref{eq:main}) and present the findings in figure \ref{fig:2sidedVaidyaplot}.

As can be seen from the left panel of figure \ref{fig:2sidedVaidyaplot}, there is very good agreement between $v_{I}$ and $v_{E}/(1-f)$. This constitutes evidence in support of formula (\ref{eq:main}). Let us make a few more comments about this figure:
\begin{itemize}
    \item As $f \rightarrow 0$, our numerical checks and analytic expression for $v_{E}(f)$ agrees with the analytical formula given in \cite{Liu:2013qca, Liu:2013iza} for AdS-RN.
    \item We note also that $v_{E}/(1-f)$ is always smaller than $v_{B}$ over the whole range of $f$. This is a striking difference from the spin-chain plot (figure \ref{fig:mut}), where $v_{E}/(1-f)$ exceeds $v_{B}$ for $f$ close enough to $1$, resulting in a ``plateau'' where $v_{I}$ saturates at large $f$. In the holographic case, we have instead that $v_{E}$ tends to $v_{B}(1-f)$ as $f \rightarrow 1$. This is checked analytically in appendix \ref{App:ve f to 1}, and is consistent with the findings in \cite{Mezei2018}.
    \item From equation (\ref{holographicf}) for $f$, we note that $f$ is not dependent on the parameter $z_{i,-}$ (this is the location of the inner horizon of the black hole if there were no infalling shell). A priori, it might be that $v_{E}$ is dependent on both $f$ and $z_{i,-}$. In CFT terms, such a dependence would mean that $v_{E}$ depends on both $f$ and another parameter characterizing the initial state, such as the pre-quench charge density. While the behavior of extremal surfaces which probe the pre-shock geometry (type O) is in principle affected by other parameters of that geometry, the limit of large regions and late times which defines $v_{E}$ eliminates dependence on pre-shock parameters besides $z_{i,+}$, as is evident in equation (\ref{eq:veAnalytic}).  Therefore $v_{E}$ is only a function of $f$, not of $z_{i,-}$.
\end{itemize}

\subsection{Beyond strips: spherical subregions}\label{subsec:sphere}

In this final subsection, we extend our holographic findings to spherical subregions instead of strips. While the use of strips is naturally adapted to initial perturbations with support on a ``wall'', the simplest shape to consider for pointwise perturbations is a sphere.\\
It is known that, for $f=0$, the thermalization of spheres of fixed sizes is considerably more complicated than for strips \cite{Liu:2013iza, Liu:2013qca, Mezei:2016zxg}: the saturation can be either continuous or discontinuous. In the case of the Vaidya background for $f=0$, the saturation for spheres is continuous when the final geometry is uncharged and is discontinuous when the final geometry is charged and sufficiently near extremality \cite{Mezei:2016zxg}. Furthermore, even when the saturation for spheres is discontinuous, the growth in the entanglement entropy before saturation is not in general linear in time \cite{Mezei:2016zxg}. When the saturation for spheres is continuous, the entanglement entropy only grows linearly at early times, then smoothly levels off and saturate at a second-order phase transition \cite{Liu:2013iza, Liu:2013qca}.

As for strips, an interesting question for spheres is how $v_{I}$ varies with $f$. We have studied this question numerically in the case where the final geometry is a charged background near extremality in 3+1 dimension. We present our findings in Figure \ref{fig:SpherevIoff} below. The blackening factor for the final geometry is taken to be $h_{f}(z) = 1 -3z^{3} + (1.4297)^{2}z^{4}$ in this figure.

\begin{figure}[h!]
\begin{center}
\includegraphics[height=0.6\columnwidth, width=0.6\columnwidth]
{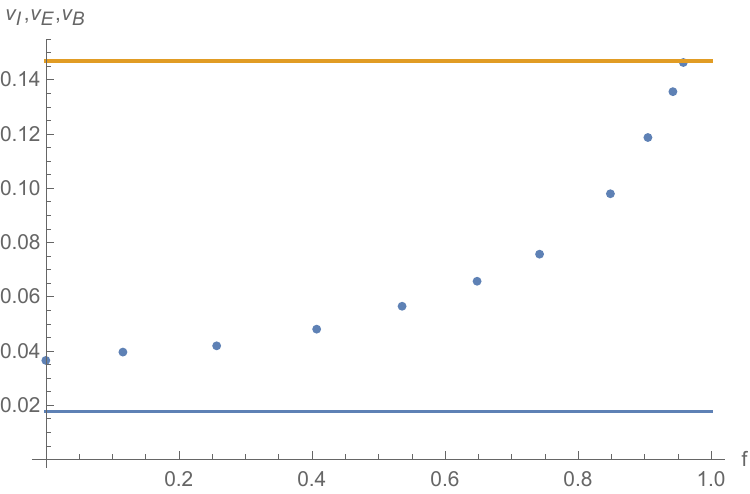}
\caption{ $v_{I}$ as a function of $f$ (blue dots). The horizontal blue line is $v_{E}$, and the horizontal orange line is $v_{B}$.}
\label{fig:SpherevIoff}
\end{center}
\end{figure}
We discuss out a few salient features of the plot above:
\begin{itemize}
    \item We see that, at $f = 0$, $v_{I}$ is approximately twice $v_{E}$. This can be understood as a consequence of the fact that, for $f=0$ and when the final geometry is near extremality, the growth in the entanglement entropy before saturation (for a large sphere of fixed size) comes close to being linear in time. Indeed, for a fixed subregion of any shape, provided that the entanglement entropy grows linearly before saturation, the saturation time is determined by:
    \begin{equation}
        t_{sat} = \frac{|A|}{v_{E}|\partial A|}
    \end{equation}
    For a sphere on a timeslice of a 2+1 dimensional field theory, we have $|A| = \pi R^{2}$ and $|\partial A| = 2\pi R$, where $R$ is the radius. The formula above becomes $t_{sat} = \frac{R}{2v_{E}}$. Also, $v_{I}$ is related to the saturation time by:
    \begin{equation}
        v_{I} = \lim_{R \rightarrow \infty} \frac{R}{t_{sat}{(R)}}
    \end{equation}
    Plugging the saturation time found above, we then find $v_{I} = 2v_{E}$, in agreement with figure \ref{fig:SpherevIoff}.
    \item There exists a critical value of $f<1$, which we will denote by $f_{c}$, at which the $v_{I}$ curve intersects the $v_{B}$ line \footnote{We thank M\'ark Mezei for first pointing out this behavior in the case of spherical boundary regions, and for useful correspondence about this case.}.  For $f_{c} < f < 1$, we have $v_{I} = v_{B}$. This is a qualitative difference from the strip case (figure \ref{fig:2sidedVaidyaplot}): in the strip case, recall that $v_{I} < v_{B}$ for all $f < 1$.\\
    When $f < f_{c}$, the smallest entanglement wedge containing the infalling particle is at a discontinuous transition. On the other hand, when $f_{c} < f < 1$, that smallest entanglement wedge is at a continuous transition.\\
    The fact that $f_{c}$ is close to $1$ is due to the fact that the final geometry is quite close to extremality. If we decrease the electric charge of the final geometry, we expect $f_{c}$ to also decrease.
\end{itemize}
Next, we comment on the validity of the Hayden-Preskill argument in section (\ref{sec:qi}) for spherical subregions. We discuss the cases of a discontinuous saturation and continuous saturation separately and explain why - in each case - the Hayden-Preskill argument remains valid:
\begin{itemize}
    \item \textbf{Discontinuous transition}. This case is similar to strips. Recall that for strips, the Hayden-Preskill argument says that $v_{I}$ is either $v_{E}/(1-f)$ (which is the speed controlling the saturation process) or $v_{B}$ (the speed controlling operator spreading), whichever is the smaller one. For spheres and for discontinuous transition, we want to carry out the same minimization. However, the speed controlling the saturation process is no longer dictated by $v_{E}$, and there is no simple closed-form formula for it. Nevertheless, because the speed controlling the saturation process is smaller than $v_{B}$ (see the portion $f < f_{c}$ of figure \ref{fig:SpherevIoff}), the Hayden-Preskill argument concludes that $v_{I}$ is the speed of the subregion at saturation.\\
    Holographically, it is indeed true that the smallest entanglement wedge containing the infalling particle is also the entanglement wedge at saturation. Thus, the Hayden-Preskill argument is verified.
    \item \textbf{Continuous transition}. This case is peculiar in that $v_{B}$ controls the saturation process, in addition to operator-spreading. To see that $v_{B}$ controls the saturation, we note that the tip of the entanglement wedge at saturation grazes the in-falling shell (by hypothesis of a continuous transition). Thus, to determine the speed of expansion of the entanglement wedge at saturation, we could pretend that the bulk spacetime were an eternal black hole, and the pre-shell geometry plays no role. We then conclude that the entanglement wedge at saturation expands at speed $v_{B}$ \footnote{We also note that, for $f=0$, the fact that $v_{B}$ controls saturation was checked explicitly in \cite{Liu:2013iza, Liu:2013qca}.}.\\
    It follows then that the minimization process discussed in section \ref{sec:qi} becomes trivial, and we find immediately that $v_{I} = v_{B}$. Holographically, the entanglement wedge at saturation is also precisely the smallest entanglement wedge containing the infalling particle. Thus, the Hayden-Preskill argument is again verified.
\end{itemize}

\section{Information transmission and decoding} 
\label{sec:transmission}

In this section, we discuss in what sense information is actually being transmitted at the speed $v_I$. What we have shown so far is that information is spreading in space at a speed $v_I$, but as discussed in the introduction, we do not expect this spreading effect to manifest as a localized motion. Instead, the information is being scrambled up into more complex degrees of freedom. However, we show that if one is willing to complete a complex decoding process, one can view the information as being transmitted at speed $v_I$. We also analyze a spatially local variant of the traversable wormhole story and show that information is again being sent at speed $v_I$, this time in a locally detectable form.

\subsection{Signalling protocol using Hayden-Preskill}

What should we mean by information transmission? If transmission is taken to mean local detectability, then barring special protected modes (e.g., hydrodynamic sound), information is simply not transmitted in a fashion that is locally detectable in chaotic systems. On the other hand, if we are allowed to act over a large region with a complex quantum operation, then information might be transmitted.

Let us say that information initially located at position $x=0$ is transmitted at speed $v_T$ if it is the largest speed $v$ such that information can be decoded by acting only on degrees of freedom with $|x| \gtrsim v t$. In other words, $v_T$ is the largest speed such that we can decode the information without access to any degree of freedom within $v_T t$ of $x=0$. We are allowed to use arbitrary degrees of freedom that are further than $v_T t$ from $x=0$.

For information traveling in localized form, say in some kind of wavepacket, the transmission speed is the wavepacket speed. If $v_T$ were higher than the wavepacket speed, then we could not get the information since it is localized in the wavepacket. If $v_T$ were less than the wavepacket speed, we could recover the information (since we can act on everything else, which includes the wavepacket), but this would be the fastest speed with this property. Hence, $v_T$ is the wavepacket speed.

The key result of this section is that, with $v_T$ defined as above, the Hayden-Preskill protocol implies that $v_T = v_I$. This is because given access to the complement, we only need a small piece of the region $|x| \leq v_I t$ to recover the information. In other words, if one controls all of a system with $|x| \geq R$, then one has access to the information after a transmission time $t_T = R/v_I$. The conventional dynamics of the system could be switched off at that time, and the information could be recovered by applying a complex quantum operation to the degrees of freedom in the region $|x| \geq R$.

\subsection{Signalling protocol using traversable wormholes}\label{sec:Wormhole}

We now turn to the setting of traversable wormholes and consider an eternal black hole rendered traversable by coupling the two boundaries with a spatially localized coupling. For previous related studies, see  \cite{Shenker:2014cwa, Gao:2016bin, Maldacena:2017axo, Almheiri:2018ijj, Caceres:2018ehr, Fu:2018oaq, Freivogel:2019lej, Fu:2019vco, Marolf:2019ojx}. More specifically, we will compute a certain squared commutator, which quantifies the degree of traversability of the wormhole. We will show that there is a time window (``the sweet spot'') where the traversability is the largest, that this sweet spot starts to appear around the scrambling time, and that the sweet spot propagates at the information speed. To discuss the simplest possible case, we consider a two-dimensional conformal field theory in which the information speed is equal to the speed of light for all $f$ and an initial state with $f=1$.

\subsubsection{A few generalities}
\begin{figure}
  \centering
    \includegraphics[width=0.7\textwidth]{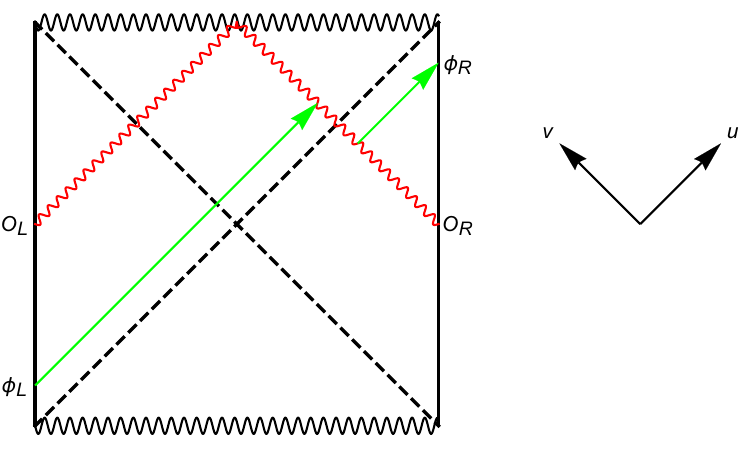}
      \caption{Setup of the traversable wormhole calculation. The coupling between the two sides gives rise to a wavefunction which has negative null energy (in red). The message (in green) undergoes a negative Shapiro delay.}
      \label{fig:wormhole}
\end{figure}

This subsection reviews the formalism of traversable wormhole as presented in \cite{Maldacena:2017axo}. Consider a generic eternal black hole. In Kruskal coordinates, the metric takes the form:
    \begin{equation}
        ds^{2} = -a{(UV)}dUdV + r^{2}{(UV)}dy^{2}
    \end{equation}
We define $a_{0} = a{(0)}$ to be the value on the horizon of the functions $a$, and denote by $r_{+}$ the horizon radius. We take the time coordinate to increase upward on both boundaries. We insert a localized coupling between the two boundaries of the form:
    \begin{equation}
        \delta H{(t_{0},x_{0})} = \frac{g}{K} \sum_{i=1}^{K} \mathcal{O}_{R}^{i}{(t_{0},x_{0})} \mathcal{O}_{L}^{i}{(t_{0},x_{0})}
    \end{equation}
where we use a large number $K$ of light fields, all of which have the same conformal dimension, to take advantage of certain simplifications. Note that the coupling is nonzero only at time $t_{0}$ and position $x_{0}$. Also, we would like to send a signal using $\phi$-quanta from the location $(t_{L}, x_{L})$ on the left boundary to the location $(t_{R}, x_{R})$ on the right boundary. We illustrate this general set-up in figure \ref{fig:wormhole}. 
Coupling the two boundaries generates negative null energy in the bulk \cite{Gao:2016bin}. As the $\phi$-quanta sail through this negative null energy, it undergoes a Shapiro time advance (rather than the usual time delay) and escapes to the other side. On the other hand, the backreaction of the $\phi$-quanta generically make the wormhole longer and therefore less traversable. Due to these two competing effects, we expect traversability to be limited to certain sweet spots.\\
We will detect traversability by studying the commutator
\begin{equation}
\langle [\phi_{L}, \phi_{R}] \rangle_{\mathcal{V}} \equiv \langle [ \phi_{L}, e^{-i\mathcal{V}} \phi_{R} e^{i\mathcal{V}} ] \rangle  
\end{equation}
where $\phi_{L,R}$ have boundary spacetime labels $\phi_{L}=\phi_{L}{(t_{L},x_{L})}$, $\phi_{R} = \phi_{R}{(t_{R},x_{R})}$, and $\mathcal{V} = \int_{t_0}^{t} dt_{1} \delta H{(t_{1},y_{0})}$. We will often not explicitly write the spacetime label for $\phi_{L,R}$ for compactness of notation. As argued in \cite{Maldacena:2017axo}, a nonzero value of the commutator above means we have managed to send information across the wormhole.\\
Assuming the operators are Hermitian, we can equivalently study the correlator:
    \begin{equation}
        C \equiv \langle e^{-i\mathcal{V}} \phi_{R} e^{i\mathcal{V}} \phi_{L} \rangle
    \end{equation}
since the original correlator is proportional to the imaginary part of $C$. Using large-$K$ factorization, we moreover have:
\begin{equation}
    C = e^{-i\langle \mathcal{V} \rangle} \tilde{C}
\end{equation}
with
\begin{equation}\label{CtildeDefinition}
    \tilde{C} = \langle \phi_{R} e^{i\mathcal{V}} \phi_{L} \rangle
\end{equation}
After some manipulations, the details of which we will skip here but which we describe in appendix \ref{app:wormhole}, we can express $\tilde{C}$ in the form: 
\begin{equation}\label{Ctilde1}
    \tilde{C} = \alpha \int dp^{U} dx p^{U} \Psi_{2}^{*}{(p^{U},x)} \Psi_{3}{(p^{U},x)} e^{i\mathcal{D}}
    \end{equation}
    with
    \begin{equation}\label{D}
        \mathcal{D} = \alpha g \int dq^{V} dy q^{V} \Psi_{1}^{*}{(q^{V},y)} \Psi_{4}{(q^{V},y)} e^{i\delta{(s,b)}}
    \end{equation}
    and
    \begin{equation}\label{alpha}
        \alpha = \frac{a_{0}^{2}}{4\pi r_{+}^{d-1}}
    \end{equation}
    \begin{equation}\label{delta}
        \delta = \frac{4\pi G_{N} a_{0}}{r_{+}^{d-3}} p^{U} q^{V} f{(|x-y|)}
    \end{equation}
    \begin{equation}\label{f}
        f{(|x - y|)} = \frac{\mu^{(d-4)/2}}{2(2\pi|x - y|)^{(d-2)/2}} e^{-\mu|x-y|}
    \end{equation}
    \begin{equation}\label{musquared}
        \mu^{2} = \frac{2\pi(d-1) r_{+}}{\beta}
    \end{equation}
    In the above, $e^{i \delta}$ is basically the scattering amplitude coming from a certain shockwave computation (as explained in appendix \ref{app:wormhole}). Furthermore, the function $\Psi_{1}$, $\dots$, $\Psi_{4}$ are the wavefunctions of the quanta, which are also the Fourier transform along the horizon of the bulk-to-boundary propagator:
    \begin{equation}\label{Psi1}
        \Psi_{1}{(p^{V},x)} = \int dU e^{ia_{0}p^{V}U/2} \langle \Phi_{O}{(U,V,x)} O_{L}{(t_{1})} \rangle |_{V=0}
    \end{equation}
    \begin{equation}\label{Psi2}
        \Psi_{2}{(p^{U},x)} = \int dV e^{ia_{0}p^{U}V/2} \langle \Phi_{\phi}{(U,V,x)} \phi_{R}{(t_{R})} \rangle |_{U=0}
    \end{equation}
    \begin{equation}\label{Psi3}
        \Psi_{3}{(p^{U},x)} = \int dV e^{ia_{0}p^{U}V/2} \langle \Phi_{\phi}{(U,V,x)} \phi_{L}{(t_{L})} \rangle |_{U=0}
    \end{equation}
    \begin{equation}\label{Psi4}
        \Psi_{4}{(p^{V},x)} = \int dU e^{ia_{0}p^{V}U/2} \langle \Phi_{O}{(U,V,x)} O_{R}{(t_{1})} \rangle |_{V=0}
    \end{equation}
    where $\Phi_O$ is the bulk field dual to $\mathcal{O}$, and $\Phi_{\phi}$ is the bulk field dual to $\phi$. We will also adopt the notation $K(U,V,x;U_{1},x_{1})$ for the bulk-to-boundary propagator between point $(U,V,x)$ in the bulk and point $(U_{1}, x_{1})$ on the boundary.

\subsubsection{Example of planar BTZ black hole}
We now specialize to the planar BTZ black hole in order to have analytic tractability. The disadvantage of specializing to this solution is that $v_{B}$ is just the speed of light in 3 dimensions. Nevertheless, this will be sufficient for our purposes.\\
The bulk-to-boundary propagator for the planar BTZ is given in Kruskal coordinates by:
\begin{equation}
    K{(U,V,x;U_{1},x_{1})} = \frac{r_{+}^{\Delta}}{2^{\Delta+1}\pi}  \left[ \frac{1+UV}{(1-UV)\cosh{(r_{+} \Delta x )} + U_{1}V - \frac{U}{U_{1}}} \right]^{\Delta}
\end{equation}
where $\Delta x = x - x_{1}$. Also, the specific forms of the wavefunctions $\Psi_{1}$, $\Psi_{2}$, $\Psi_{3}$ and $\Psi_{4}$ are written down in the appendix (see equations (\ref{psi1})- (\ref{psi4})). Plugging those wavefunctions into (\ref{Ctilde1}) and (\ref{D}), and performing the $q^V$ integral exactly gives:
\begin{eqnarray}\label{TripleIntegral}
\tilde{C} &=& \alpha_{\psi} e^{r_{+}\Dpsi(t_{L}-t_{R})} \int_{0}^{\infty} dp^{U} \int_{-\infty}^{\infty} dx  (p^{U})^{2\Dpsi-1} e^{2ip^{U}[\cosh{(r_{+}(x-x_{R}))} e^{-r_{+}t_{R}} + \cosh{(r_{+}(x-x_{L}))} e^{r_{+}t_{L}}]} \nonumber \\
&\times& \exp{\left\{ ig e^{i\pi\DO}\Gamma{(2\DO)}\alpha_{\mathcal{O}}   \int_{-\infty}^{\infty} dy \left[ 4\cosh{(r_{+}(y-x_{0}))} \cosh{(r_{+}t_{0})} + 8\pi p^{U} G_{N} e^{-r_{+}|x-y|} \right]^{-2\DO} \right\}} \nonumber
\end{eqnarray}
where we have defined
\begin{equation}
    \alpha_{\psi} = \alpha \frac{r_{+}^{2\Dpsi}}{4} \frac{e^{-i\pi \Dpsi}}{(\Gamma{(\Dpsi)})^{2}}    
\end{equation}
\begin{equation}        \alpha_{\mathcal{O}} = \alpha \frac{r_{+}^{2\DO}}{4} \frac{e^{-i\pi \DO}}{(\Gamma{(\DO)})^{2}}    
    \end{equation}
Let us now focus on the innermost integral (over $y$).  With $u=r_+(y-x_0)$ this integral becomes:
    \begin{equation}
    \frac{1}{r_+}\int_{-\infty}^{\infty} du \left[ 4\cosh{(u)} \cosh{(r_{+}t_{0})} + 8\pi  G_{N} p^{U}e^{-|r_{+}(x-x_0)-u|} \right]^{-2\DO}
    \end{equation}
This is a difficult integral to evaluate analytically, so we seek an approximate form. Numerical investigation reveals that, to a good approximation, the integral above only depends on $p^{U}$ and $x$ through the combination $p^{U}e^{-r_+|x-x_0|}$, and is well approximated by 
    \begin{equation}\begin{split}
    \frac{2^{-(2\Delta_O+1)}\Gamma(\Delta_O)^2}{r_+ \Gamma(2\Delta_O)}
    \left(\cosh{(r_+t_0)} + 4\pi G_{N} p^{U}e^{-|r_{+}(x-x_0)|}\right)^{-2\DO}
    \end{split}\end{equation}
    With this approximation and a shift of the integration parameter $x\rightarrow x+x_0$, $\tilde{C}$ becomes
    \begin{eqnarray}\label{ApproximateCtilde}
        \tilde{C} &=& \alpha_{\psi} e^{r_+\Dpsi(t_{L}-t_{R})} \int_{0}^{\infty} dp^{U} \int_{-\infty}^{\infty} dx  (p^{U})^{2\Dpsi-1} e^{2ip^{U}[\cosh{(r_+(x-\Delta x_{R}))} e^{-r_+ t_{R}} + \cosh{(r_+(x-\Delta x_{L}))} e^{r_+t_{L}}]} \nonumber \\
        &\times& \exp{\left[ igH(\DO)\left(1+\frac{4\pi G_{N}p^U}{\cosh{(r_+ t_0)}} e^{-r_+|x|}\right)^{-2\DO} \right]}
    \end{eqnarray}
    where we have introduced the notation $\Delta x_R=x_R-x_0$, $\Delta x_L=x_L-x_0$, and defined
    \begin{equation}\label{eq:HDO}
        H{(\DO)} =\frac{\alpha}{4}\frac{r_+^{2\DO-1}}{2^{2\DO+1}}(\cosh{(r_+ t_0)})^{-2\Delta_0}
    \end{equation} 
To facilitate the identification of the ``sweet spot" for communication we change variables to $u= A p^U e^{r_+ x}$ and $v= B p^U e^{-r_+ x}$, with
\begin{equation}\begin{split}\label{eq:AB}
        A&=(e^{-r_+(\Delta x_{R}+t_{R})} + e^{-r_+(\Delta x_{L}-t_{L})})\\
        B&=(e^{r_+(\Delta x_{R}-t_{R})} + e^{r_+(\Delta x_{L}+t_{L})})
\end{split}\end{equation}
The integral can then be brought to the form:
    \begin{equation}\begin{split}\label{Ctilde}
    \tilde{C} =&\ \frac{\alpha_{\psi}}{2r_+} e^{r_+\Dpsi(t_{L}-t_{R})} (AB)^{-\Delta_\psi}\\
    &\times \bigg(
    \int_{0}^\infty dv v^{\Dpsi-1}\exp\left[i v +igH(\Delta_\mathcal{O})\left(1+\frac{4\pi G_N}{\cosh{(r_+ t_0)}} v/B\right)^{-2\Delta_\mathcal{O}}\right] \int_{v\frac{A}{B}}^\infty du u^{\Dpsi-1}\exp(i u)\\
    &+ \left(A\leftrightarrow B\right)\bigg)\\
    \end{split}\end{equation}

We write the inner integrals in terms of the exponential integral function:  $\int_a^\infty du\ u^{\Delta-1}\exp(i u)= a^{\Delta} E_{1-\Delta}\left(-i a\right)$.
Now we recall that $C$ differs from $\tilde{C}$ by a phase factor ($C=e^{-i g H(\DO)}\tilde{C}$), which can be computed using the same method as the computation of $\tilde{C}$ presented here (see Appendix \ref{app:wormhole}). And so we arrive at the following expression for $C$ to be integrated numerically:
    \begin{equation}\begin{split}\label{C}
    C&=\ \frac{\alpha_{\psi}}{2r_+} e^{r_+\Dpsi(t_{L}-t_{R})}(AB)^{-\Dpsi}\times
    \int_{0}^\infty dv \hspace{1mm}v^{2\Dpsi-1}\\
    &\bigg[\left(\frac{A}{B}\right)^{\Dpsi}
    E_{1-\Dpsi}\left(-i v \frac{A}{B}\right)\exp\left(i v -igH(\Delta_\mathcal{O})\left(1-\left(1+\frac{4\pi G_N}{\cosh{(r_+ t_0)}} \frac{v}{B}\right)^{-2\Delta_\mathcal{O}}\right)\right)\\&+ \left(A\leftrightarrow B\right)\bigg]\\
    \end{split}\end{equation}
    
\subsubsection{The light-cone}
\begin{figure}\label{fig:ButterflyConePlots}
\centering
\begin{subfigure}[t]{0.8\textwidth}
\centering
\includegraphics[width=\textwidth]{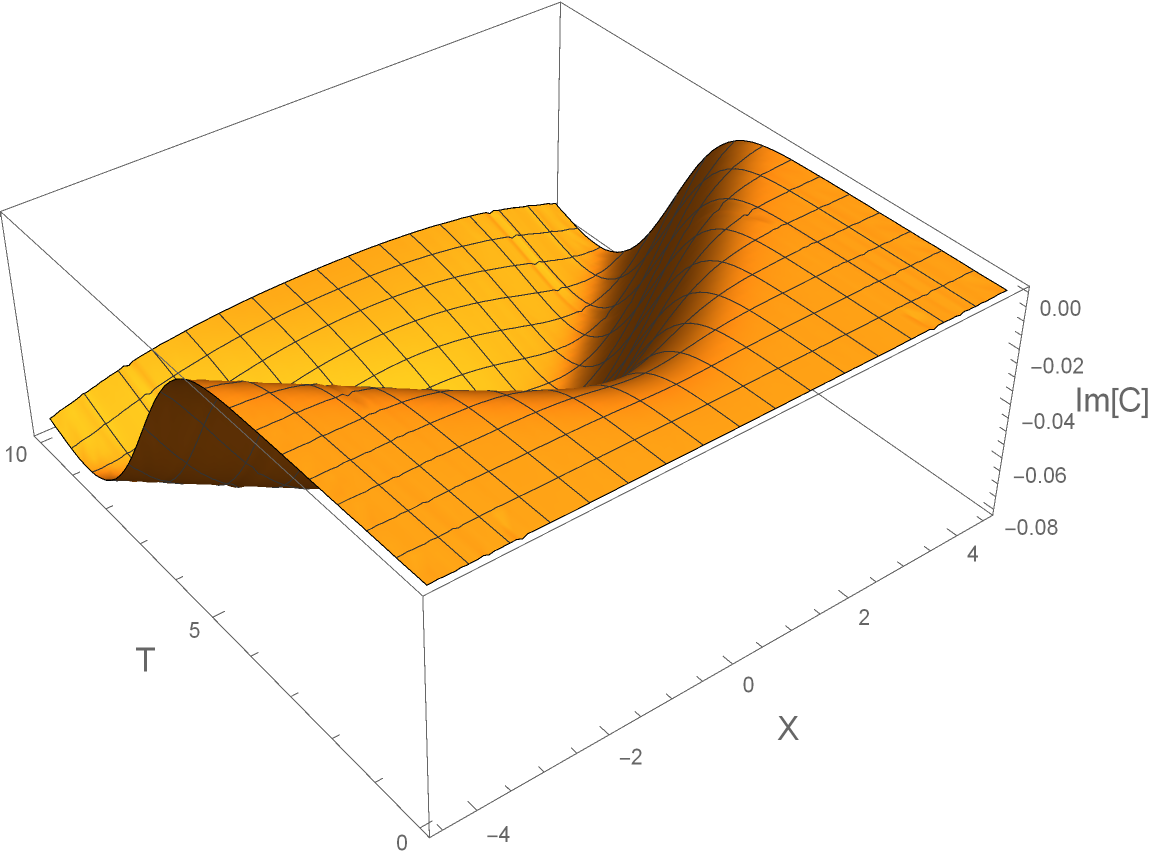}
\end{subfigure}
\begin{subfigure}[t]{0.8\textwidth}
\centering
\includegraphics[width=\textwidth]{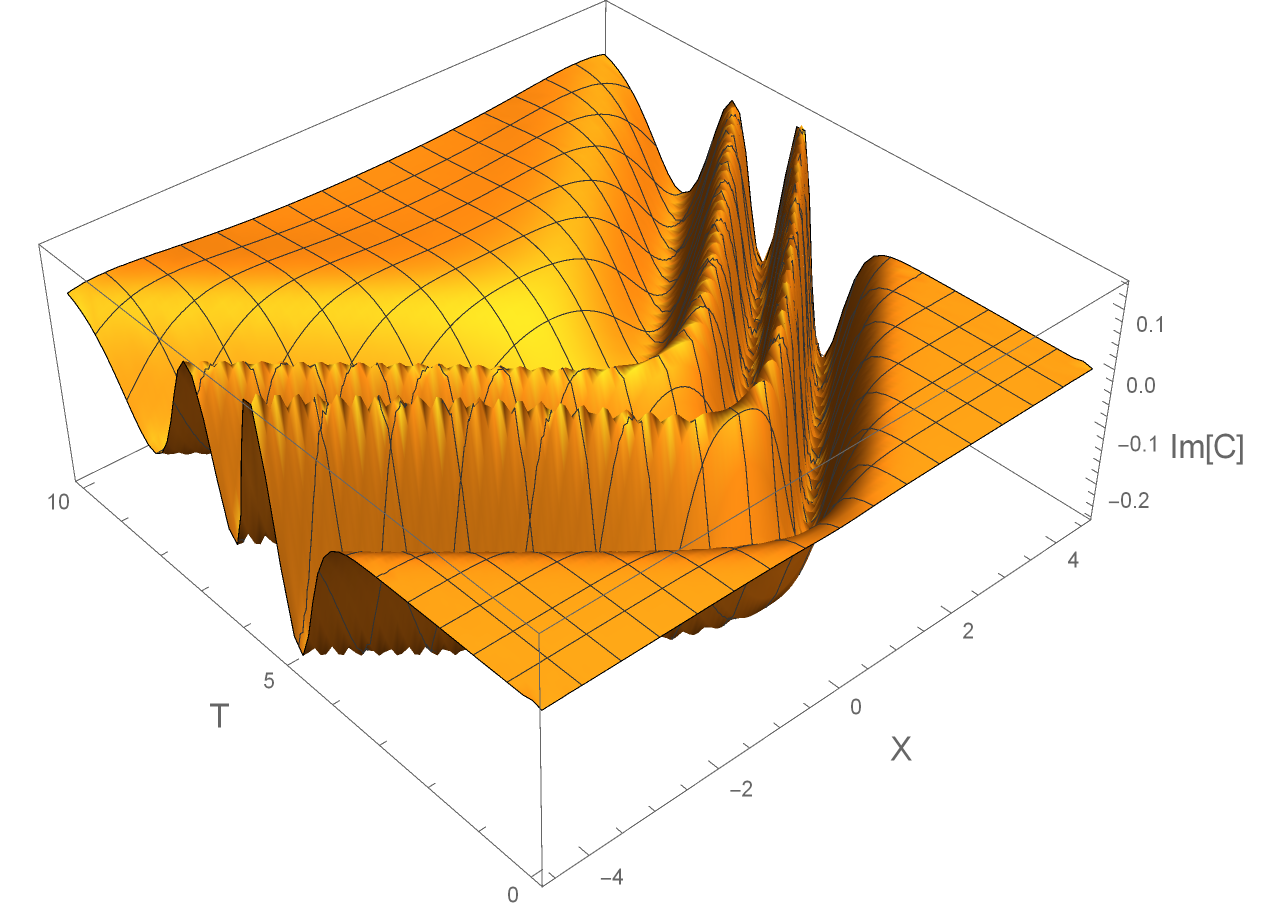} 
\end{subfigure}
 \caption{Plot of the imaginary part of $C$ as a function of $T$ and $X$, for two choices of parameters. Both plots use $\DO = \Dpsi = 0.7$, and $t_{scr} = 5$.  The upper figure has $g=30$ while the lower figure has $g=300$.}
\label{LightConePlots}
\end{figure}

Equation (\ref{C}) gives us $C$ as a function of $x_{L}$, $x_{R}$, $t_{L}$ and $t_{R}$. Note that the dependence on those four boundary spacetime coordinates is entirely through $A$ and $B$. In order to see the ``sweet spot'' of traversability from (\ref{C}), we now set $t_0=0$, $-t_{L} = t_{R} \equiv T$ and $\Delta x_{L} =\Delta x_{R} = X$, and study the function $C(T,X)$. 
\begin{equation}\begin{split}
C(T,X)=&\ \frac{\alpha_{\psi}}{2r_+}e^{i\frac{\pi}{2}\Dpsi} \times
\int_{0}^\infty dv \hspace{1mm} v^{2\Dpsi-1}\\
&\bigg[
\Gamma\left(\Dpsi,-ive^{-2r_+ X}\right)\exp\left(i v -igH(\Delta_\mathcal{O})\left(1-\left(1+2\pi G_N v e^{r_+(-X+T)} \right)^{-2\Delta_\mathcal{O}}\right)\right)\\&+ \left(X\leftrightarrow -X\right)\bigg]\\
\end{split}\end{equation}
We present 3-dimensional plots in Figure \ref{LightConePlots} for the imaginary part of this function over $X$ and $T$ for various values of $g$.\\
As can be seen from the plot, the quantity $C$ is essentially zero outside a lightcone centered at the bilocal coupling (i.e., at $X=0$). Furthermore, $C$ has the largest magnitude near the edge of the lightcone, an effect which is more pronounced for larger coupling $g$. This indicates that there is indeed a sweet spot of traversability that propagates that the speed of light. 

It is possible to give an analytical argument for the presence of this lightcone, as follows. We first define the variable $\delta$ by $X= T-\delta$.  We then have
\begin{equation}
    A = 2 \exp\left(r_+(-2 T+\delta)\right)
\end{equation} 
\begin{equation}
    B=2 \exp(-r_+\delta)
\end{equation}
In the limit of large $X$, $T$, at fixed $\delta$, $A\ll1$. Looking back to equation (\ref{Ctilde}) for $\tilde{C}$, the lower limits on the two nested integrals are sent to $v\frac{A}{B}\rightarrow 0$ and $u\frac{B}{A}\rightarrow \infty$.  This allows us to ignore the latter integral and approximate (\ref{Ctilde}) as
\begin{eqnarray}
    \tilde{C} &\approx& \frac{\alpha_{\psi}}{2r_+} \int_{0}^{\infty} dv v^{\Dpsi-1} \exp{[iv + igH(\DO) (1+4\pi G_{N}v/B)^{-2\DO}]} \int_{0}^{\infty} du u^{\Dpsi-1} e^{iu} \nonumber
\end{eqnarray}
In the form above, $\tilde{C}$ depends on $X$ and $T$ only through the combination $G_N /B = \frac{1}{2} G_N e^{r_+\delta}$, and this combination is of order one when $T = X + \frac{L^2}{r_+}\ln \frac{1}{G_N}$ (momentarily restoring the AdS scale).\\
This shows not only that the sweet spot propagates at the speed of light but also that the lightcone starts to open a time of the order the scrambling time after the bilocal coupling.\\
We end by noting that our computation of the correlator $C$ is expected to become unreliable when the ``message'' is sent at very large $T$. This is because the scattering amplitude $e^{i\delta}$ that we used is valid in the elastic eikonal approximation (where $G_{N} s$ is of order unity, with $s$ being the Mandelstam parameter of the scattering). At very large $T$, the relative boost between the $\phi$-quanta and the $O$-quanta is very large, and $G_{N} s$ (where $s$ is the Mandelstam parameter of the collision) is no longer of order unity. In that regime, inelastic effects are expected to become important \cite{Shenker:2014cwa}. 

\section{Discussion}
\label{sec:discussion}

In this paper, we presented a comprehensive theory of quantum information propagation in generic chaotic systems. Our general theory uses information-theoretic ideas, specifically the Hayden-Preskill protocol, to track the flow of quantum information in spacetime. We also gave strong evidence that the theory applies to both quantum spin chains and holographic quantum field theories. Much of the analysis focused on one-dimensional systems or simple strip regions, but the generalization to higher dimensions and more complex shapes is straightforward. One simply needs to track the entanglement of various regions and the growth of local operators.

In the case of holography, our results gave new insights into the way dynamics in the dual holographic geometry are related to chaos and strong coupling speed limits on the boundary. In particular, the motion of the HRT surface was very sharply constrained by general features of local quantum chaotic systems.

We also gave a preliminary discussion of information transmission by showing that information can, in some sense, be transmitted at the speed $v_I$. However, the resulting output is highly scrambled and may be difficult to decode. In the special case of $f=1$, we were able to set up a fine-tuned initial state where there was a locally detectable signal that propagated at speed $v_B$. This is a spatially local analog of the recently studied traversable wormhole setup.

More generally, one can verify that the initially localized information is indeed fully scrambled in the sense that any piece of the scrambled output, when sent to the complement, is sufficient to recover the entanglement. In AdS/CFT, this is a statement about the behavior of HRT surfaces when thermal scale fragments of a region are removed.

There are numerous questions and directions. Here are a few, in no particular order:
\begin{itemize}
    \item One interesting direction is to further quantify the properties of the quench experiment discussed here when viewed as a communication channel. In particular, in the traversable wormhole setup, a local signal is detectable, but it is not clear to what extent this local signal corresponds to a high fidelity transmission of quantum information.
    
    \item Recent work has raised the possibility of multiple butterfly speeds depending on different ``thermal regulators'' used to define the out-of-time-order correlator~\cite{Liao_2018_OTOC_Metal_Regulator,RomeroBermudez_2019_OTOC_Regulator}. Given our work, it is natural to identify the true chaos speed as the one related to information propagation. If the butterfly speed does depend on the regulator, one may speculate that the symmetric regulator is the one related to information propagation, but this remains open.
    
    \item Another important question is the effect of fluctuations. Recent work has argued that the growth of operators generically yields a wavefront that broadens with time~\cite{Xu_2018_Scrambling_Fluctuations}. As long as the broadening is sub-ballistic, the identification of the information speed should be unaffected in the asymptotic limit. However, fluctuations will make the transition from recoverability in $A$ to recoverability in $A^c$ rather non-sharp. It would be extremely interesting to understand these effects in holography, especially their impact on entanglement wedge reconstruction and the traversable wormhole setup.
    
    \item It would also be interesting to extend our analysis to cases without translation symmetry, including random disorder and quasiperiodic potentials, where new slow dynamics can emerge~\cite{Sahu2018}. 
    
    \item Throughout this work, we have focussed on large subregions in the CFT. It is interesting to understand the features associated with thermalization of small subregions, such as the ones discussed in \cite{Caceres:2012em, Kundu:2016cgh}, in terms of spreading and propagation of quantum information. It is also interesting to construct spin chain models that exhibit the same phenomenology.
    
    \item The quench protocol that defines our information speed provides a generalization of Hayden-Preskill where the necessary entanglement is not performed. As we discussed extensively, it becomes necessary first to generate nearly maximal entanglement. As such, our protocol may be adaptable to diagnose the difference between weak and strong scrambling recently discussed by Shor~\cite{Shor_2018_Scrambling_BlackHole}.
    
    \item Hydrodynamic sound (and related protected modes, like Goldstone modes) provides an exception to the general expectation that local excitations cannot propagate in a strongly chaotic system. The chaotic spin chain we studied does not exhibit such a mode at infinite temperature, and we did not include such modes in our holographic analysis. Presumably, as a hydrodynamic excitation, sound is not able to coherently transport quantum information; however, it is important to better understand how the existence of a hydrodynamic sound mode interacts with the setup presented here. 
    
    \item Finally, it is important to devise methods to probe the information speed in experiment. Of course, the relevant entropies can be extracted from a full tomography of the state, but that approach is not scalable to large size. The traversable wormhole setup is one approach that is scalable given access to a sufficiently sophisticated quantum simulator or fault-tolerant quantum computer. In that context, it is important to understand the effects of fluctuations and whether the setup can be extended to $f<1$. We are also studying other schemes to track measures of information spreading based on wavefunction overlaps.

\end{itemize}

\begin{acknowledgments}
	This material is based upon work supported by the National Science Foundation under Grant Number PHY-1620610. PN is supported by NSF grant PHY-1708139. The work BGS is supported in part by the Simons Foundation via the It From Qubit collaboration and in part by the U.S. Department of Energy, Office of Science, Office of High Energy Physics QuantISED Award de-sc0019380. This material is also based in part upon work of XS supported by the U.S. Department of Energy, Office of Science, Advanced Scientific Computing Research Quantum Algorithms Teams program, QOALAS collaboration. We thank Elena Caceres for an interesting discussion on holographic thermalization, and M\'ark Mezei for insights regarding the case of spherical boundary regions.   
\end{acknowledgments}

\appendix

\section{Holographic Derivation of $v_E$, $v_I$ for general $f$}
\label{App:ve}

In this appendix, we adapt the procedure of Liu and Suh (\cite{Liu:2013qca},\cite{Liu:2013iza}) to a derivation of entanglement velocity for the case of nonzero initial entropy fraction $f$.  In \cite{Liu:2013qca}, the time dependence of entanglement entropy and related variables under a thermal quench are studied holographically through the behavior of bulk extremal surfaces associated with fixed boundary regions.  The corresponding bulk geometries are vacuum AdS in the past, transitioning to a black hole spacetime in the future via a uniform null shell stress tensor (or ``shockwave"). Several universal features of the time dependence are identified. Among these is the entanglement velocity $v_E$, which quantifies a period of linear growth of entanglement entropy associated with large boundary regions after the quench and before equilibration:
\[\Delta S_\Sigma = s_{\text{eq}}A_\Sigma v_E t+O(1)\]
Here $\Delta S_\Sigma$ is the regularized entanglement entropy of a boundary region $\Sigma$, computed as the difference between the minimal area (divided by $4G_N$) among bulk extremal surfaces homologous to $\Sigma$ and the minimal area (divided by $4G_N$) associated with the same boundary region in empty AdS.  $A_\Sigma$ is the boundary ``area" (or volume) of $\Sigma$, and $s_\text{eq}$ is the equilibrium entropy density.

In this section, we consider as initial states not vacuum AdS, but another black hole metric corresponding to a nonzero initial entropy density.  The entanglement fraction $f$ is the ratio of initial entropy density to final entropy density, and we seek the analogue of the above expression for general $f$.   We specialize to the case of ``strips" as boundary regions. The extent to which results generalize to other large boundary regions and shapes is an interesting question which we mostly leave to future work (though see the related discussion in section IX.B of \cite{Liu:2013qca}).

\subsection{Metric and notation}
We consider a special case of AdS-Vaidya style metrics, given by 
\begin{equation}\begin{split}\label{eq:shockmetric}
ds^2 = \frac{L^2}{z^2}\left(-h(v,z)dv^2-2dv dz +d\vec{x}\cdot d\vec{x}\right)\\
h(v,z)= h_i(z)+\Theta(v)(h_f(z)-h_i(z))
\end{split}\end{equation}
where $\Theta(v)$ is a Heaviside step function, so $h_{i}(z)$ and $h_{f}(z)$ describe the black hole geometry in the initial ($v<0$) and final ($v>0$) regions, respectively, separated by the shock at $v=0$. Either $h_i(z)$ or $h_f(z)$ could describe a static, planar black hole in AdS (with $h(0)=1$), with metrics in terms of static time $t$ given by
\[ds^2=\frac{L^2}{z^2}\left(-h(z)dt^2+h(z)^{-1}dz^2+d\vec{x}\cdot\vec{x}\right)\]
Both may or may not have multiple horizons, but since the outermost horizon in each region will be of special interest to us, we will designate these as $z_{i,+}$ and $z_{f,+}$. We assume these static geometries can be glued together with a sensible shockwave profile, which at least requires that $z_{f,+}<z_{i,+}$ for a positive energy injection.

In addition to the instances above, throughout this appendix we will use subscripts $i$ and $f$ to denote ``initial" and ``final" regions. For these metrics, we also define the equilibrium densities $a_{\text{eq}}=\left(\frac{L}{z_{f,+}}\right)^{n}\frac{1}{4G_N}$,  $s_\text{eq}=\left(\frac{L}{z_{f,+}}\right)^{d-1}\frac{1}{4G_N}$, and the initial entanglement fraction $f=\left(\frac{z_{f,+}}{z_{i,+}}\right)^{d-1}$.

\subsection{Boundary strips}
The simplest boundary regions to consider are generalized ``strips" of finite width in the $x_1$ direction (-$R$ to $R$), infinite extent in the $x_2$ through $x_n$ direction, and $x_{n+1}, ... , x_{d-1}$ set to zero. The case immediately of interest to us is $n=d-1$, and the corresponding boundary observable is an entanglement entropy.  The formulas reported here do not assume this however.

For the associated extremal surfaces, symmetry dictates that we can parametrize the surface $v$ and $z$ coordinates as a function of $x_1\rightarrow x$ and pull out an overall (infinite) factor of the area in the $x_2$ through $x_n$ transverse space, which we denote $A_{n-1}$.  The area functional is then:

\begin{equation}\begin{split}\label{eq:AreaFunctional}
A&=A_{n-1}L^n\int z^{-n}\sqrt{Q}\hspace{1mm}dx
\hspace{2mm}\\
\text{ with }\\
Q&=Q(v,z)=\left(1-h(v,z)(v')^2-2v' z'\right)
\end{split}\end{equation}
Here and in the rest of this appendix we will often suppress the arguments of functions such as $Q$, $h$,...etc for compactness.  The z and v equations of motion are given by
\begin{equation*}\begin{split}
\partial_z Q-2 n z^{-1}Q &= z^n\sqrt{Q}\partial_x\left(\frac{\partial_{z'}Q}{z^n\sqrt{Q}}\right)\\
\partial_v Q &= z^n\sqrt{Q}\partial_x\left(\frac{\partial_{v'}Q}{z^n\sqrt{Q}}\right)
\end{split}\end{equation*}

\subsection{Symmetries and Conserved Quantities}
The lack of explicit $x$ dependence results in a constant of motion on extremal surfaces:
\begin{equation}\begin{split}\label{eq:J}
J=z^n\sqrt{Q}=z^n\sqrt{1-h(v,z)(v')^2-2v' z'}
\end{split}\end{equation}
This allows us to simplify both equations of motion:
\begin{equation}\begin{split}\label{eq:zEOM}
\partial_z Q-2nz^{-1}Q=\partial_x \partial_{z'}Q\\
\end{split}\end{equation}
\begin{equation}\begin{split}\label{eq:vEOM}
\partial_v Q=\partial_x \partial_{v'}Q\\
\end{split}\end{equation}
In both regions $v>0$ and $v<0$, the metric (\ref{eq:shockmetric}) is independent of $v$.  Setting $\partial_v h_{i,f}=0$ in the $v$ equation (\ref{eq:vEOM}) gives a quantity which carries a constant value on either side of the shock:
\begin{equation}\begin{split}\label{eq:E}
E&= z'+h v'\\
&=E_i + \Theta(v)\left(E_f-E_i\right)
\end{split}\end{equation}
Using this to eliminate $v'$ from (\ref{eq:J}) gives a useful differential equation for $z$ which can then be inverted and integrated to find $x(z)$.
\begin{equation}\begin{split}\label{eq:zeom}
z'=\mp\sqrt{h\left(\frac{J^2}{z^{2n}}-1\right)+E^2}
\end{split}\end{equation}
In conjunction with (\ref{eq:E}) this can be used to find an equation governing $v(z)$:
\begin{equation}\label{eq:v(z)}
\frac{dv}{dz}=
-h^{-1}\left(1\pm\frac{E}{\sqrt{h\left(\frac{J^{2}}{z^{2n}}-1\right)+E^2}}\right)    
\end{equation}

Finally, employing these in the area functional (\ref{eq:AreaFunctional}) yields
\begin{equation}\begin{split}\label{eq:Area}
\frac{dA}{dx}&=A_{n-1}L^n J z^{-2n}\hspace{1mm}\\
\end{split}\end{equation}

The solutions to these equations can be uniquely specified, up to surfaces equivalent through symmetry, by the coordinate position of the ``tip", or the deepest point of the extremal surface: $z=z_t$ and $v=v_t$ (we also use a subscript ``t" for other quantities evaluated at this ``tip").  At this point, smoothness of the surface dictates that $v'_t=z'_t=0$, implying through (\ref{eq:E}) that $E_i=0$, so we may use $E=\Theta(v)E_f$. Equation (\ref{eq:zeom}) then indicates that $J=z_t^{n}$.

\subsection{Junction conditions}
It still remains to determine the value of $E_f$ for a given $z_t, v_t$.  To do this we examine the effect of the shock itself on the extremal surface. The function $h(v,z)$ is not independent of $v$ at the shock, so we must return to equations (\ref{eq:zEOM}) and (\ref{eq:vEOM}).  Let the subscript ``$c$" denote quantities evaluated at the position of shock crossing, and suppose the extremal surface meets the shock ($v=0$) at $\hspace{1mm} z=z_c, \hspace{1mm} x=x_c$.  Integrating across the shock from $x=x_c-\epsilon$ to $x=x_c+\epsilon$, and sending $\epsilon\rightarrow 0$ gives give junction conditions for $v'$ and $z'$, though the former is trivial.
\begin{equation}\begin{split}\label{eq:junctions}
v'_{f,c}&=v'_{i,c}\\
z'_{f,c}&=\frac{z'_{i,c}}{2}\left(1+\frac{h_f(z_c)}{h_i(z_c)}\right)
\end{split}\end{equation}
Considering the implications for $J$ and $E$ on either side of the shell confirms the constancy of $J$ and determines the value of $E_f$.
\begin{equation}\label{eq:E+}
E_f=\frac{z_{i,c}'}{2}\left(1-\frac{h_f(z_c)}{h_i(z_c)}\right)
=-\frac{1}{2}\sqrt{h_i(z_c)\left(\frac{z_t^{2n}}{z_c^{2n}}-1\right)}
\left(1-\frac{h_f(z_c)}{h_i(z_c)}\right)
\end{equation}
where we have used equation (\ref{eq:zeom}) and restricted to surfaces with $z'(x)<0$ before the shock (surfaces initially heading toward the boundary). 

\subsection{Governing equations}

Everywhere off the shock itself, the governing equations which we will integrate to find $R$, $t$, and $A$ become:

\begin{equation}\begin{split}\label{eq:mainEOMs}
\frac{dx}{dz}&=\frac{\mp 1}{\sqrt{H(z)}}\\
\frac{dv}{dz}&=\frac{-1}{h(z)}\left(1\pm \frac{E}{\sqrt{H(z)}}\right)\\
\frac{dA}{dz}&=A_{n-1}(L z_t)^n\frac{\mp 1}{z^{2n}\sqrt{H(z)}}
\end{split}\end{equation}
with
\[\begin{split}
H(z)=h(z)\left(\frac{z^{2n}_t}{z^{2n}}-1\right)+E^2
\end{split}\]
The top signs are used if $z(x)$ is a monotonically decreasing function, but in regions where $z$ increases with $x$ the lower sign is used.
Note that because $h(z)$ and $E$ effectively take on different meaning for $v>0$ and $v<0$ according to
\[\begin{split}
&h(z)=h_i(z)+\Theta(v)\left(h_f(z)-h_i(z)\right)\\
&E=\Theta(v)E_f
\end{split}\]
$H(z)$ can also be thought of as piecewise function.
\[\begin{split}
H(z)&=H_i(z)+\Theta(v)\left(H_f(z)-H_i(z)\right)\\
H_i(z)&=h_i(z)\left(\frac{z^{2n}_t}{z^{2n}}-1\right)\\
H_f(z)&=h_f(z)\left(\frac{z^{2n}_t}{z^{2n}}-1\right)+E_f\\
\end{split}\]

Thus far, it has been convenient to specify surfaces by their tip coordinates $(z_t,v_t)$. However, from here it will be convenient to think of the same family of surfaces as parametrized by $z_t$ and $z_c$ (of course, the middle of equations (\ref{eq:mainEOMs}) can be integrated on the pre-shock side to find $z_c(v_t,z_t)$). 

\subsection{Surfaces at late time $t$}
We are interested in areas of extremal surfaces associated with late times $t$ and large boundary width $R$. With the boundary time set to track time elapsed after the shock, we let $t=v$ on the boundary.  The boundary time associated with an extremal surface is then obtained by integrating the middle of equations (\ref{eq:mainEOMs}) only over the region outside the shock.  This indicates that large $t$ may be reached when $H_{f}(z)$ is near zero for a substantial portion of the integration range.  This expectation is confirmed by more detailed analysis in section $VII$ of \cite{Liu:2013qca}, which we cursorily summarize here.

It will be useful to examine $H_f(z)$ as a multivariate function of $z, z_t, z_c$:
\begin{equation}\begin{split}\label{eq:Hf}
H_f(z)=H_f(z,z_t,z_c)= h_f(z)\left(\frac{z^{2n}_t}{z^{2n}}-1\right)+E^2_f(z_t,z_c)\\
E_f^2(z_t,z_c)=\frac{1}{4}h_i(z_c)\left(\frac{z_t^{2n}}{z_c^{2n}}-1\right)\left(1-\frac{h_f(z_c)}{h_i(z_c)}\right)^2
\end{split}\end{equation}
We now define several critical quantities: $z_m(z_t), z_c^*(z_t), z_s$, and $z_t^{(s)}$.  Although we will usually leave the $z_t$ dependence implicit, the first two can be thought of as functions of the tip depth $z_t$.  The latter two are fixed values in a given spacetime.  

The following discussion applies to $z_t>z_{f,+}$.  First define $z_m$ as the location of the local minimum of $H_f(z)$ between and $z_{f,+}$ and $\text{min}(z_t, z_{f,-})$, $z_{f,-}$ being the second smallest positive real root of $h_{f}(z)$, if it exists.

%
%
Next define $z_c^*(z_t)=z_c^*$ to be the smallest value of $z_c>z_{f,+}$ which results in $H_f(z_m,z_t,z_c^*)$=0.  A surface which crosses the shock close to this value results a trajectory which lingers very near $z_m$ for a long time while the area and width grow (lingering where $H(z)$ is small).  Crossing the shock at $z_c^*$ itself results in a surface which approaches $z_m$ and remains there indefinitely as $v\rightarrow \infty$.

Even among those surfaces which do reach the boundary, some do so non-monotonically (in terms of $z(x)$).  To understand this, note that if $\frac{h_f(z_c)}{h_i(z_c)}<-1$, the latter of equations (\ref{eq:junctions}) indicates that as $z_c$ crosses this value it results in a sign change for $z'$ as the surface crosses the shock.  Define $z_s$ as the smallest $z$ greater than $z_{f,+}$ such that $\frac{h_f(z_s)}{h_i(z_s)}=-1$. 

Lastly, considering $z_m$ as a function of $z_t$, there might occur a special value of $z_t$, denoted $z_t^{(s)}$, such that $z_m(z_t^{(s)})=z_s$.  It also so happens that at such a point $z_c^*(z_t^{(s)})$ is also equal to $z_s$, so at this point we have:
\[z_m(z_t^{(s)})=z_c^*(z_t^{(s)})=z_s\]

We now restrict to considering surfaces with $z_c$ close to but less than $z_c^*$, setting through $z_c=z_c^*(1-\epsilon)$ with $\epsilon<<1$.  One of two basic scenarious then occurs: if $z_t<z_t^{(s)}$, then $z(x)$ decreases monotonically toward the boundary while if $z_t>z_t^{(s)}$, the surface initially falls inward before turning around and heading to the boundary. In the monotonic case, from equations (\ref{eq:mainEOMs}) the boundary parameters are given by
\begin{gather}\label{eq:monotonic}
R=\int_0^{z_c}\frac{dz}{\sqrt{H_f{(z)}}}+\int_{z_c}^{z_t}\frac{dz}{\sqrt{H_i{(z)}}}\\
t=\int_0^{z_c}\frac{dz}{h_f(z)}\left(\frac{E_f}{\sqrt{H_f(z)}}+1\right)\\
\frac{A}{A_{n-1}}\left(L z_t\right)^{-n}=\left(\int_{z_{\text{cutoff}}}^{z_c}\frac{dz}{z^{2n}\sqrt{H_f(z)}}
+\int_{z_c}^{z_t}\frac{dz}{z^{2n}\sqrt{H_i(z)}}\right)
\end{gather}
while for the non-monotonic case we have
\begin{gather}\label{eq:nonmonotonic}
R=\left(\int_0^{z_r}dz+\int_{z_c}^{z_r}dz\right)\frac{1}{\sqrt{H_f{(z)}}}+\int_{z_c}^{z_t}dz\frac{1}{\sqrt{H_i{(z)}}}\\
t=\left(\int_0^{z_r}dz+\int_{z_c}^{z_r}dz\right)\left(\frac{1}{h_f(z)}\left(\frac{E_f}{\sqrt{H_f(z)}}+1\right)\right)\\
\frac{A}{A_{n-1}}\left(L z_t\right)^{-n}=\left(\left(\int_{z_\text{cutoff}}^{z_r}dz+\int_{z_c}^{z_r}dz\right)\frac{1}{z^{2n}\sqrt{H_f(z)}}
+\int_{z_c}^{z_t}dz\frac{1}{z^{2n}\sqrt{H_i(z)}}\right)
\end{gather}
where $z_r$ is the root of $H(z)$ which is below $z_m$.  

For fixed $z_t$, as $\epsilon\rightarrow0$ each of the above integrals is dominated by the part of the surface approaches and lingers near $z_m$.  To approximate the contribution from this region, first expand $H_f(z)$ around $z=z_m$ and also in small $\epsilon$. Using the fact that $H(z_m)=0$ when $\epsilon=0$, we have
\begin{equation}\begin{split}\label{eq:Happrox}
H_f(z)&\approx b\epsilon+H^f_2(z-z_m)^2
\end{split}\end{equation}
with $b=- z^*_c \frac{d E^2}{dz_c}\bigg|_{z_c=z_c^*}$ and $H^f_2=\frac{1}{2}\frac{d^2H_f}{dz^2}\bigg|_{z=z_m}$.

Integrating across this crucial region, to leading order in small $\epsilon$, $R, t$, and $A$ receive contributes as
\begin{gather}\begin{split}\label{eq:epsilon contribution}
t&=-\frac{E(z_c^*)}{h_f(z_m)\sqrt{H^f_2}}\log{\epsilon}+...\\
R&=-\frac{1}{\sqrt{H^f_2}}\log{\epsilon}+...\\
\frac{\Delta A}{A_{n-1}}\left(L z_t\right)^{-n}&=-\frac{1}{z^{2n}_m\sqrt{H^f_2}}\log{\epsilon}+...
\end{split}\end{gather}

In the last line, $\Delta A= A-A_{\text{vac}}$ and we have subtracted off the area of a region of the same $R$ in empty $AdS$, which eliminates the divergent part of the area but otherwise contributes negligibly.  At this point, the ellipses contain all the other parts of the integrals.  The remaining contributions between the boundary and $z_c$ are at most $O(z_{f,+})$.  

We now proceed to estimate the size of the contributions to $A$ and $R$ from the region prior to the shock, the integrals from $z_c$ to $z_t$.  

\begin{gather}\label{eq:preshockAR}
R\supset\int_{z_c}^{z_t}dz\frac{1}{\sqrt{H_i{(z)}}}\\
\frac{A}{A_{n-1}}\left(L z_t\right)^{-n}\supset\int_{z_c}^{z_t}dz\frac{1}{z^{2n}\sqrt{H_i(z)}}
\end{gather}

\subsection{Limit $f\rightarrow 0$}\label{subsection:small f}
For pre-shock geometries with relatively small mass, $z_{i,+}>>z_{f,+}$ (or in the limit of empty AdS), we can push $z_t>>z_{f,+}, z^*_{c}$.  In this case, $R$ receives a contribution linear in $z_t$, while area and time receive negligible contributions.  Thus by sending $z_t$ arbitrarily large we can reach a regime where $R>>t>>z_{f,+}$, and equations \ref{eq:epsilon contribution} then indicate that 
\[\begin{split}\Delta A&=t \times\frac{z_t^n}{z_m^{2n}}\frac{h_f(z_m)}{E(z_c^*)}L^n A_{n-1}+...\\
&= t \times z_m^{-n} \sqrt{-h_f(z_m)}\left(1-\frac{z_m^{2n}}{z_t^{2n}}\right)^{-\frac{1}{2}}L^n A_{n-1}+...
\end{split}\]
where in the last equality we used (\ref{eq:Hf}) and the definition of $z_c^*$ as a root of $H_f(z)$.  For general $n$ we let $a_{eq}=\frac{L^n}{z_{f,+}^n}$ and for $z_t>>z_m$ this becomes
\begin{equation}\begin{split}\label{eq:define vE}
\Delta A&\approx a_{\text{eq}}A_{n-1}v_n t+...\\
v_n&=\frac{z_{f,+}^{n}}{z_m^{n}} \sqrt{-h_f(z_m)}
\end{split}\end{equation}

In the case of $n=d-1$, using $s_{\text{eq}}=\frac{1}{4G}\frac{L^{d-1}}{z_{f,+}^{d-1}}$, we can rewrite this as
\begin{equation}\begin{split}\label{eq:Liu Suh vE}
\Delta S&\approx s_{\text{eq}}A_{d-2}v_E t+...\\
v_E&=v_{d-1}
\end{split}\end{equation}

This is the result reported by Liu and Suh.   It is crucial that for large $z_t$, $z_m(z_t)$ is a very slow-changing function of $z_t$, or that for fixed $R$, $z_t$ remains essentially fixed as $t$ increases (in practice for $f= 0$ it is the former that happens, and in this regime $z_m(z_t)$ is often well approximated by its value in the $z_t\rightarrow \infty$ limit).  In either case, the prefactor gives a constant velocity, and this is a ``linear growth regime" for fixed $R$.

\subsection{Finite $f$}
In the case that there is a horizon $z_{i,+}\sim z_{f,+}$, $z_t$ cannot be pushed arbitrarily large.  To reach an analogous large $R$ limit, $z_t$ is pushed extremely close to the outermost pre-shock horizon.  Let $z_t=z_{i,+}(1-\delta)$. As $\delta\rightarrow 0$, the dominant pre-shock contributions to $A$ and $R$ then come from the region where $H_i(z)=h_i(z)\left(\frac{z_t^{2n}}{z^{2n}}-1\right)$ is smallest.  Note this function has a local minimum between $z_t$ and $z_{i,+}$ which we will denote $z_M$.  Expanding around $z_M$, to leading order in small $\delta$ we have
\begin{equation}\begin{split}\label{eq:Hmapprox}
H_i(z)&\approx H_i(z_M)+(z_M-z)^2 H_2^i\\
H_2^i &= \frac{1}{2}H_i''(z_M)
\end{split}\end{equation}
Note that, to lowest order in $\delta$, the term $H_i(z_M)$ is second order in $\delta$. Adding the contributions from integrals (\ref{eq:preshockAR}) near the upper limit, we find that to leading order in the small $\epsilon$, small $\delta$ expansion
\begin{gather}\begin{split}
t&=-\frac{E(z_c^*)}{h_f(z_m)\sqrt{H^f_2}}\log{\epsilon}+...\\
R&=-\frac{1}{\sqrt{H^f_2}}\log{\epsilon}-\frac{1}{\sqrt{H^i_2}}\log{\delta}+...\\
\frac{\Delta A}{A_{n-1}}\left(L z_{i,+}\right)^{-n}&=-\frac{1}{z^{2n}_m\sqrt{H^f_2}}\log{\epsilon}-\frac{1}{z_{i,+}^{2n}\sqrt{H^i_2}}\log{\delta}+...
\end{split}\end{gather}

Here we have used the fact that, as $\delta \rightarrow 0$, $z_t$ and $z_M$ both tend to $z_{i,+}$ (recall that $z_M$ is sandwiched between $z_t$ and $z_{i,+}$). These can be rearranged to find that in the large $R$, large $t$ limit, there is a linear growth regime governed by
\begin{equation}\begin{split}\label{eq:vn}
\Delta A = a_{\text{eq}}A_{n-1}\left(t\times v_n + \frac{z_{f,+}^n}{z_{i,+}^n}R+...\right)\\
v_n=\frac{z^n_{f,+}}{z^n_m}\sqrt{-h_f(z_m)}\left(1-\frac{z^{2n}_m}{z_{i,+}^{2n}}\right)^{\frac{1}{2}}
\end{split}\end{equation}
where the final manipulations are directly analogous to those of section \ref{subsection:small f}.  Rearranging slightly and specializing to $n=d-1$ we find the entanglement velocity to be
\begin{equation}\label{eq:ve}
v_E=\frac{z^{d-1}_{f,+}}{z^{d-1}_{i,+}}\sqrt{-h_f(z_m)\left(\frac{z^{2(d-1)}_{i,+}}{z_{m}^{2(d-1)}}-1\right)}
\end{equation}
Recall that $z_m$ in this expression is defined as the minimum with respect to $z$ of the function $h_f(z)\left(\frac{z_{i,+}^{2n}}{z^{2n}}-1\right)$ between $z_{f,+}$ and $z_{i,+}$.  We find the expression (\ref{eq:ve}) to be substantiated by purely numerical investigations into the linear growth regime for fixed boundary regions.

\subsection{Limit $f\rightarrow1$}\label{App:ve f to 1}
In general spacetimes and dimensions, analytic expressions for $z_m$ are hard to come by, but considering the large $f$ limit allows some insight.  As the pre-shock outer horizon is pushed to the post-shock outer horizon (so the initial entanglement fraction $f\rightarrow 1$), the minimum is sandwiched halfway between these horizons as long as the $h_f''(x_m)\ne 0$.  Letting
\[z_{i,+}= z_{f,+}(1+x)\]
\[z_m=z_{f,+}\left(1+\frac{1}{2}x+...\right)\]
equation (\ref{eq:ve}) becomes, to leading order in small $x$,  
\[\begin{split}
v_n&=x\sqrt{-\frac{1}{2}n z_{f,+} h_f'(z_{f,+})}\\
\end{split}\]
Or, with $n=d-1$ and trading $x$ for $f=\left(\frac{z_{f,+}}{z_{i,+}}\right)^{d-1}=1-\left(d-1\right)x+...$, we have
\[\begin{split}
v_e&\underset{f\rightarrow1}{=}\left(1-f\right)\sqrt{ \frac{2 \pi z_{f,+}T }{d-1}}+...=\left(1-f\right)v_b+...
\end{split}\]

\subsection{Saturation time and $v_I$}\label{subsec:sattimeandvI}
Return now to equation (\ref{eq:vn}):
\[\Delta A = a_{\text{eq}}A_{n-1}\left(t\times v_n + \frac{z^n_{f,+}}{z^n_{i,+}}R\right)\]
This tracks $\Delta A$ for a \textit{minimal} extremal surface only up to the saturation time, when the area is equal to its equilibrium value.  For larger $t$ type ``E" surfaces, those which do not probe the pre-shock geometry, take over.  We can approximate this saturation time $t_{sat}$ for $R,t>>z_{f,+}$ by setting $\Delta A$ to it's equilibrium value: $\Delta A_{eq}=a_{\text{eq}}A_{n-1}R$. This gives
\begin{equation}
t_{\text{sat}}= \frac{(1-\frac{z^n_{f,+}}{z^n_{i,+}})}{v_n}R
\end{equation}

In the case of $n=d-1$, turning this expression around to give the half width of the boundary region which has just reached entanglement saturation.
\begin{equation}
R_{sat}=\frac{v_E(f)}{(1-f)}t
\end{equation}
At late times this gives the smallest HRT surface to contains the particle, and we have
\begin{equation}
v_I=\frac{v_E(f)}{(1-f)}
\end{equation}

\subsection{Numerical results for strips for $f \neq 0$: a closer look}
\begin{figure}[h!]
\includegraphics[height=0.45\columnwidth, width=0.45\columnwidth]
{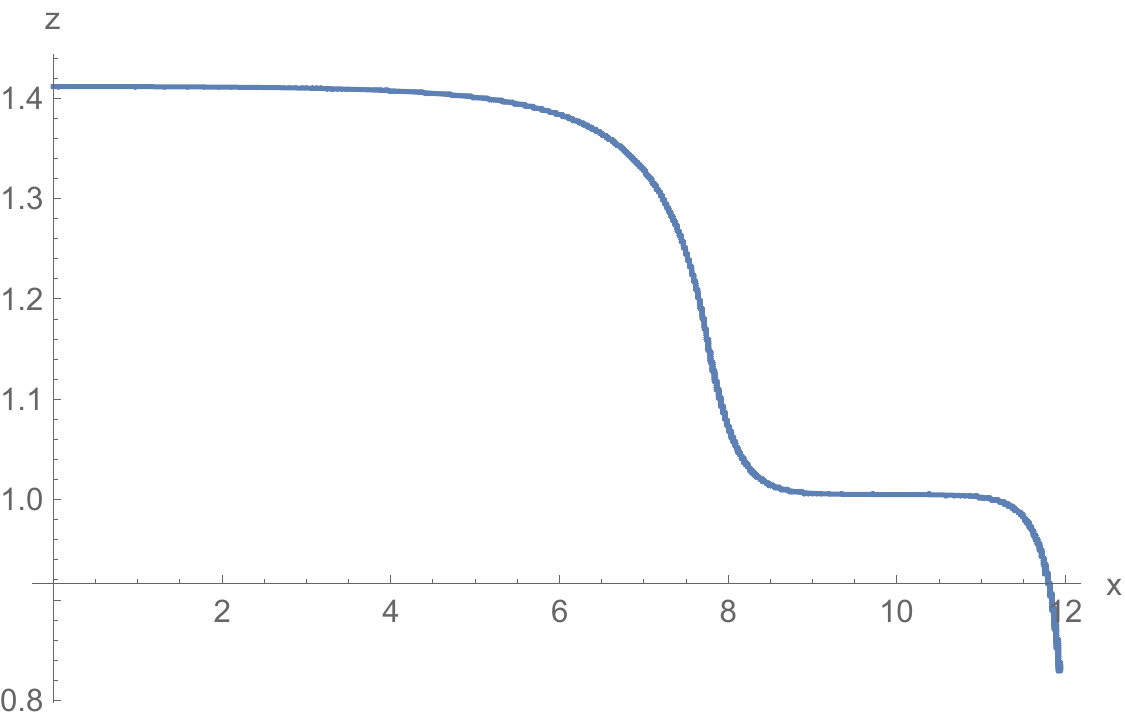}
\includegraphics[height=0.45\columnwidth, width=0.45\columnwidth]
{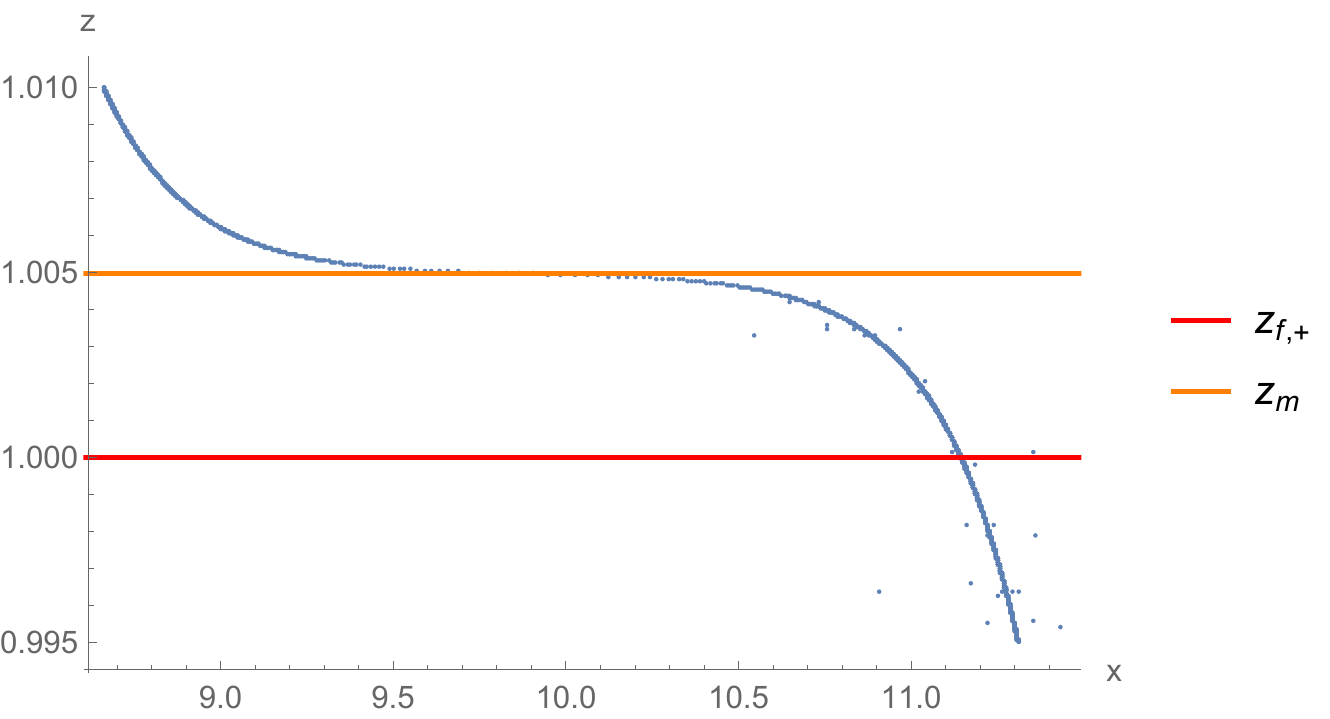}\\
\includegraphics[height=0.45\columnwidth, width=0.45\columnwidth]
{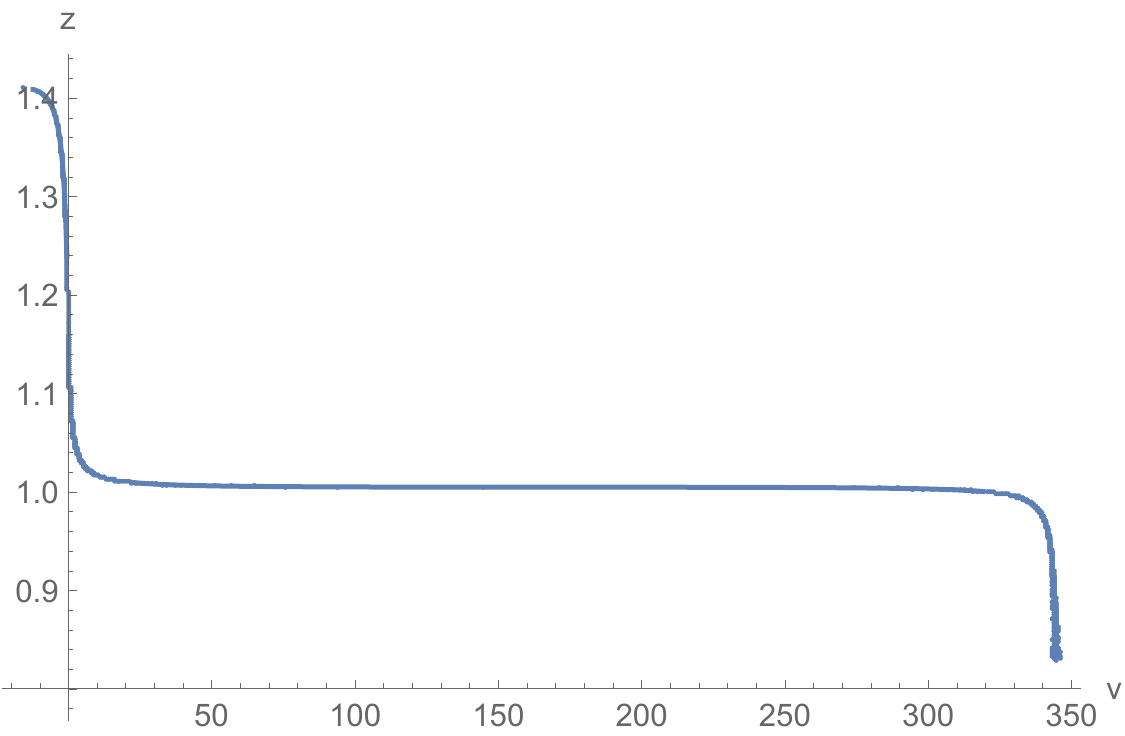}
\includegraphics[height=0.45\columnwidth, width=0.45\columnwidth]
{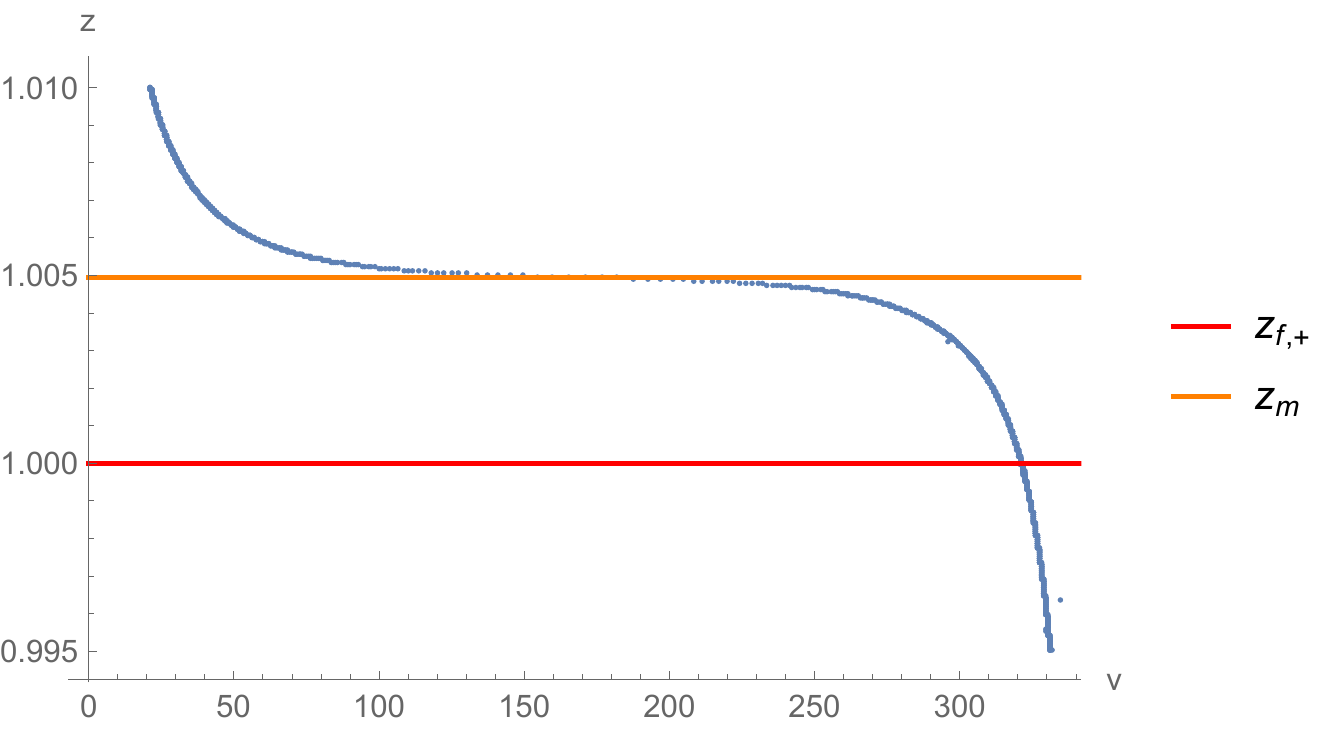}
\caption{ Upper left: Plot of $z(x)$. Upper right: Plot of $z(x)$ zoomed into the coordinate range $z_{f,+} \leq z \leq z_{m}$. Lower left: Plot of $z(v)$. Lower right: Plot of $z(v)$ zoomed into the coordinate range $z_{f,+} \leq z \leq z_{m}$. We only plotted over positive values of $x$ for both the upper left and lower left panels; the plot for negative values of $x$ can be obtained by reflection symmetry about $x=0$.} 
\label{fig:HRTplot}
\end{figure}

In this subsection, we present a few numerical results that serve as intermediate steps in the production of Figure \ref{fig:2sidedVaidyaplot} and provide some intuition for the behavior of the relevant surfaces. First, we plot the shape of a sample HRT surface in the large-width, late-time regime in Figure (\ref{fig:HRTplot}). More precisely, we plot the two functions $z(x)$ and $z(v)$ describing such an HRT surface, for the case $f=0.5$.\\
The plot of $z(x)$ (upper left panel) has two plateaux: an upper plateau and a lower plateau. These plateaux grow in size as the tip of the HRT surface approaches the apparent horizon $z_{i,+}$, and account for most of the width of the strip in the large-width regime. The upper plateau is at $z = z_{t}$, and the lower plateau occurs at $z=z_{m}$. From the upper right panel, we can also see that the slope of $z(x)$ steepens as soon as we move away from $z = z_{m}$ toward the horizon $z_{f,+}$. This suggests that the HRT surface does not linger in the coordinate range $z_{m} < z < z_{f,i}$; the surface simply passes briefly through that range on its way to the boundary.\\
The plot of $z(v)$ (lower left panel) only has one plateau. That plateau can be seen to occur at $z = z_{m}$ (lower right panel), and this plateau grows in size as the $z$ value upon crossing the shock, $z_{c}$ tends to a critical value we denote $z_{c}*$. Thus, this plateau is associated with the late-time regime. As can be seen from the lower right panel also, the slope of $z(x)$ steepens away from $z_{m}$.

\begin{figure}[h!]
\begin{center}
\includegraphics[height=0.45\columnwidth, width=0.6\columnwidth]
{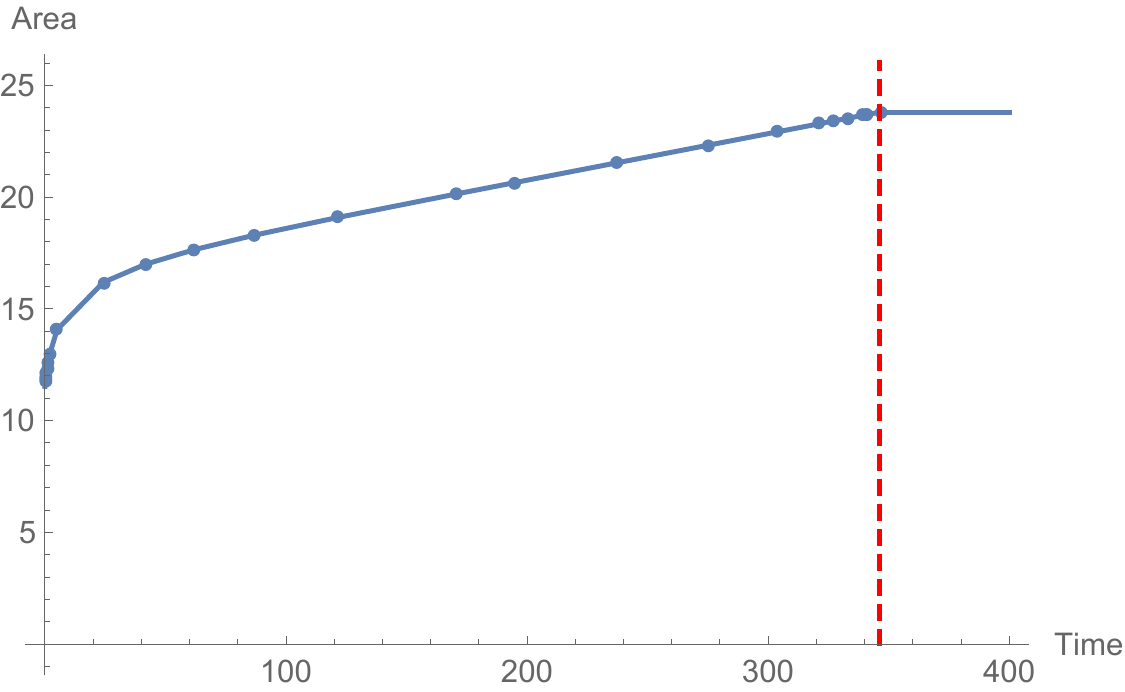}
\end{center}
\caption{ Plot of the area of the HRT surface versus boundary time, with the half-width kept fixed at the same value as in the previous plot. The dashed red vertical line is the saturation time.}
\label{fig:AreavsTime}
\end{figure}

Next, we present in figure \ref{fig:AreavsTime} the plot of the area of the HRT surface as a function of boundary time, with the half-width kept fixed at the same half-width as the sample HRT surface in figure \ref{fig:HRTplot}. As can be seen from the plot, there is a linear growth regime that persists all the way to saturation, and saturation happens at a first-order phase transition. By measuring the slope of the linear growth regime at the moment of saturation, we obtained the green dots on the right panel of figure \ref{fig:2sidedVaidyaplot} (the ones labelled ``numerical $v_{E}(f)$'' on that panel).

\begin{figure}[h!]
\includegraphics[height=0.45\columnwidth, width=0.45\columnwidth]
{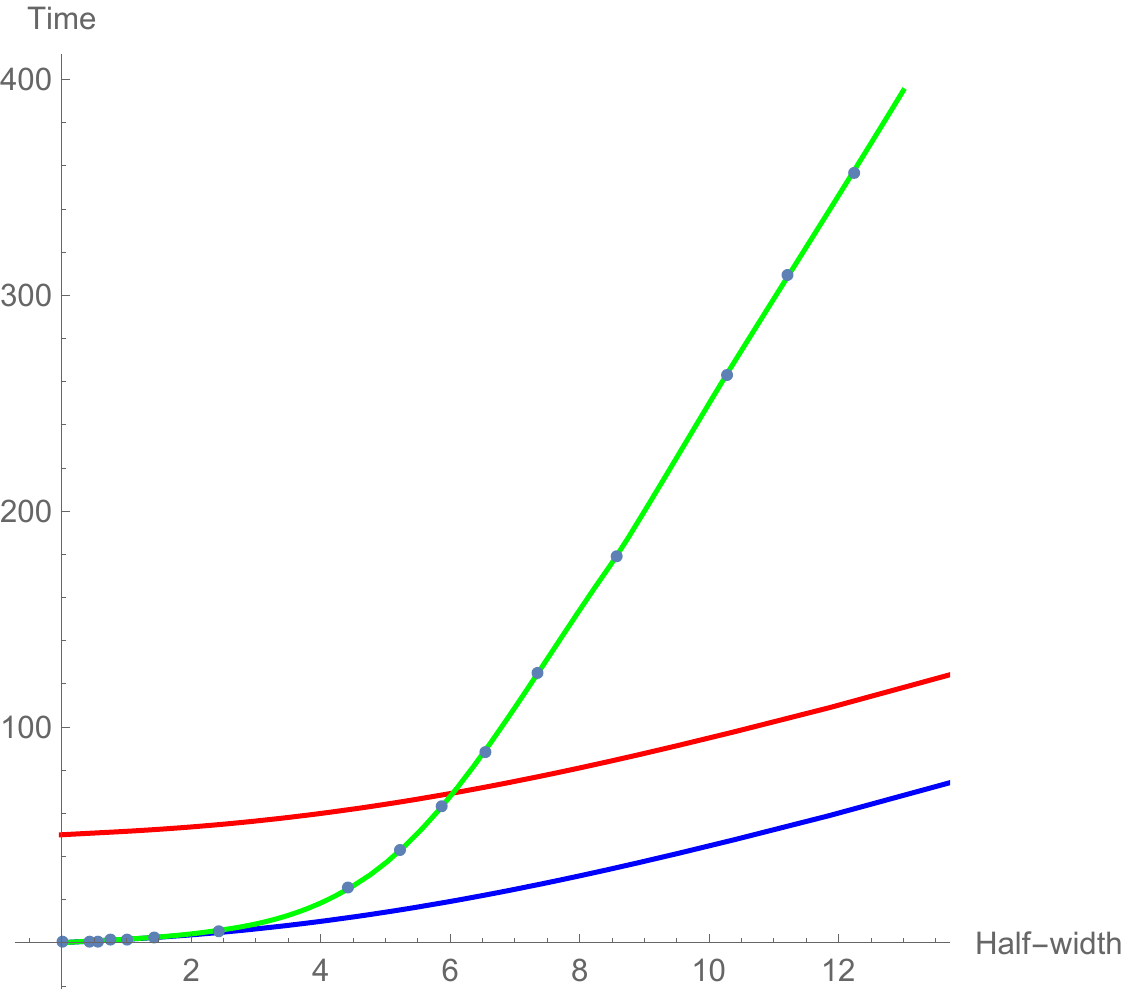}
\includegraphics[height=0.45\columnwidth, width=0.45\columnwidth]
{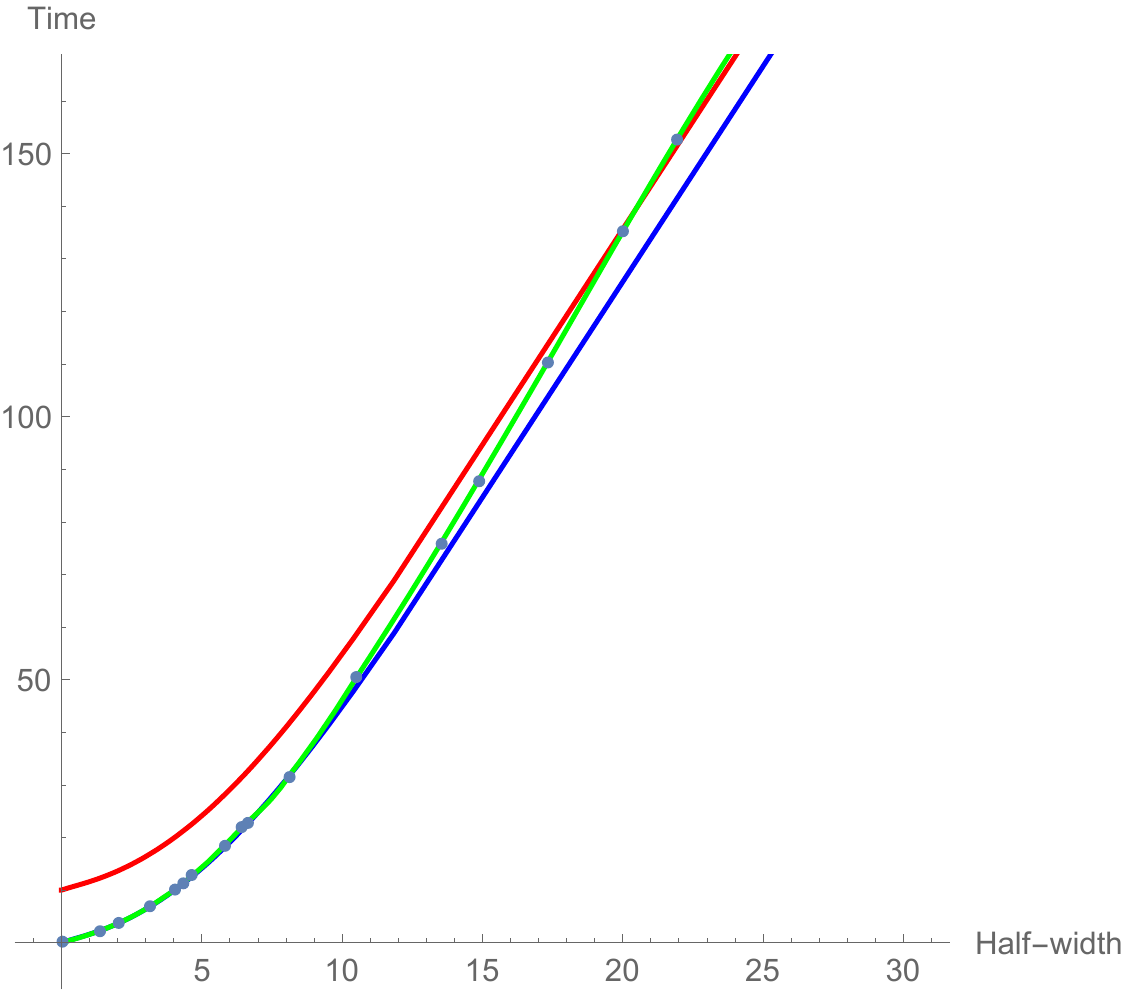}
\caption{ Left panel: Phase diagram for the case $f=0.5$. Right panel: Phase diagram for the case $f=0.99$.}
\label{fig:2sidedPhaseDiagrams}
\end{figure}

Finally, we show in figure \ref{fig:2sidedPhaseDiagrams} the ``phase diagrams'' for two nonzero values of $f$ (namely, $f=0.5$ and $f=0.99$). These phase diagrams have the same general structure as the one for $f=0$ in figure \ref{1sidedPhaseDiagram}, and the three curves (red, green and blue) have the same meaning as in figure \ref{1sidedPhaseDiagram}. The fact that the phase diagram for $f>0$ is qualitatively the same as for $f=0$ means in particular that the saturation of the entanglement entropy is a first-order phase transition for $f>0$. We also note that as $f \rightarrow 1$, the slope of the green curve approaches the slope of the other two curves. This indicates that $v_{I}$ (which is the slope of the green curve) approaches $v_{B}$ (the slope of the two other curves) in this limit. By measuring the slope of the green curves in these phase diagrams, we obtained the green dots on the left panel of figure \ref{fig:2sidedVaidyaplot} (the ones labelled ``numerical $v_{I}(f)$'' on that panel).

\section{Details of traversable wormhole calculation}\label{app:wormhole}
In this appendix, we fill in a few intermediate steps in the derivation of the quantities $\tilde{C}$ and $C$ in section \ref{sec:Wormhole}. We emphasize that the treatment here is only a slight twist of the treatments found in \cite{Almheiri:2018ijj,Maldacena:2017axo}. 

\subsection{Derivation of $\tilde{C}$}
First, we discuss the steps involved in the derivation of equations (\ref{Ctilde1})-(\ref{Psi4}) for $\tilde{C}$. Starting with the definition (\ref{CtildeDefinition}) of $\tilde{C}$, let us expand the exponential in that definition in series:
\begin{equation}\label{tildeCseries}
    \tilde{C} = \sum_{n} \frac{(ig)^n}{n!} \left \langle \phi_{R} \left( \frac{1}{K} \sum_{i=1}^{K} O_{R}^{i} O_{L}^{i} \right)^{n} \phi_{L} \right \rangle
\end{equation}
We now focus on computing the term with $n=1$ and some particular value of $i$ (from 1 to $K$), i.e. the 4-point function $\langle \phi_{R} O_{R}O_{L} \phi_{L} \rangle$.\\
We note that this 4-point function can be written as the overlap $\langle \psi' | \psi \rangle$ between an in-state $|\psi \rangle$ and an out-state $|\psi' \rangle$:
    \begin{equation}
        |\psi\rangle = O_{R} \phi_{L} |TFD\rangle
    \end{equation}
    \begin{equation}
        |\psi'\rangle = O_{L} \phi_{R} |TFD\rangle
    \end{equation}
Due to the exponential gravitational blueshift, there is a large relative boost between the $O$-wavefunction and the $\phi$-wavefunction. For such a high energy scattering process, it is convenient to use a basis of the Hilbert space labeled by the longitudinal momentum $p^{U}$ (or $p^{V}$) and the transverse position $x$. The wavefunctions for the four sources $O_{L,R}$, $\phi_{L,R}$ are then functions of $p^{U}$, $p^{V}$ and $x$, and in fact, they are the Fourier transform of the bulk-to-boundary propagators along the horizon. Explicitly:
    \begin{equation}
        \Psi_{O_L}{(p^{V},x)} = \int dU e^{ia_{0}p^{V}U/2} \langle \Phi_{O}{(U,V,x)} O_{L}{(t_{0},x_{0})} \rangle |_{V=0}
    \end{equation}
    \begin{equation}
        \Psi_{O_R}{(p^{V},x)} = \int dU e^{ia_{0}p^{V}U/2} \langle \Phi_{O}{(U,V,x)} O_{R}{(t_{0},x_{0})} \rangle |_{V=0}
    \end{equation}
    \begin{equation}
        \Psi_{\phi_L}{(p^{U},x)} = \int dV e^{ia_{0}p^{U}V/2} \langle \Phi_{\phi}{(U,V,x)} \phi_{L}{(t_{L},x_{L})} \rangle |_{U=0}
    \end{equation}
    \begin{equation}
        \Psi_{\phi_R}{(p^{U},x)} = \int dV e^{ia_{0}p^{U}V/2} \langle \Phi_{\phi}{(U,V,x)} \phi_{R}{(t_{R}, x_{R})} \rangle |_{U=0}
    \end{equation}
The 4 wavefunctions above are precisely equations (\ref{Psi1})-(\ref{Psi4}) in section \ref{sec:Wormhole}. In that subsection, to avoid notational overload, we use the subscripts $1, 2, 3, 4$ for $O_{L}$, $\phi_{R}$, $\phi_{L}$ and $O_{R}$ respectively.\\
The in-state and out-state can be expressed in terms of the wavefunctions above as:
    \begin{equation}
        |\psi\rangle = \int \Psi_{3} {(p^{U},x_{3})} \Psi_{4} {(p^{V},x_{4})} | p^{U} x_{3}; p^{V} x_{4} \rangle_{in}
    \end{equation}
    \begin{equation}
        |\psi' \rangle = \int \Psi_{1} {(p^{V},x_{1})} \Psi_{2} {(p^{U}, x_{2})} | p^{V} x_{1}; p^{U} x_{2} \rangle_{out}
    \end{equation}
The 4-point function becomes:
    \begin{equation}
        \langle \psi' | \psi \rangle = \int  \Psi_{1}^{*}{(q^{V},x_{1})} \Psi_{2}^{*}{(q^{U},x_{2})} \Psi_{3}{(p^{U},x_{3})} \Psi_{4}{(p^{V},x_{4})} {}_{out} \langle q^{V} x_{1}; q^{U} x_{2} | p^{U} x_{3} ; p^{V} x_{4} \rangle_{in}
    \end{equation}
Due to the high energy of the scattering process, the amplitude is basically diagonal in the basis above (see for example \cite{Shenker:2014cwa}):
    \begin{equation}
        | p^{V}, x_{1}; p^{U},x_{2} \rangle_{out} \approx e^{-i\delta(s,b)} | p^{V},x_{1}; p^{U},x_{2} \rangle_{in} + |\chi\rangle
    \end{equation}
with $\delta$ some phase, and $|\chi\rangle$ is a state orthogonal to all in-states. In addition, the high energy of the scattering implies that the leading contribution to the amplitude comes from gravitational interaction, and we can take $e^{i \delta}$ to come from a gravitational shockwave. The expression (\ref{delta}) given in section (\ref{sec:Wormhole}) is precisely the amplitude coming from a shockwave computation. For more details about that shockwave amplitude, see \cite{Shenker:2014cwa} and references therein.\\
Using the facts above, combined with the normalization of single-particle states:
    \begin{equation}
        \langle p^{V},x | q^{V},y \rangle = \frac{a_{0}^{2}p^{V}}{4\pi r_{+}} \delta{(p^{V} - q^{V})} \delta{(x-y)}
    \end{equation}
    we find the inner product
    \begin{equation}
        {}_{out}\langle q^{V},x_{1}; q^{U},x_{2} | p^{U}, x_{3}; p^{V},x_{4} \rangle_{in}= e^{i\delta} \left( \frac{a_{0}^{2}}{4\pi r_{+}} \right)^{2} p^{U}p^{V} \delta{(p^{U}-q^{U})} \delta{(p^{V}-q^{V})}\delta{(x_{1}-x_{4})} \delta{(x_{2}-x_{3})}
    \end{equation}
Hence, the 4-point function takes the form:
    \begin{equation}
        \langle \phi_{R} O_{L} O_{R} \phi_{L} \rangle = \alpha \int dp^{U} dx p^{U} \psi_{2}^{*}(p^{U},x) \psi_{3}(p^{U},x) \left[ \alpha \int dp^{V} dy p^{V} \psi_{1}^{*}(p^{V}, y) \psi_{4}(p^{V}, y) e^{i \delta(s)} \right]
    \end{equation}
    with
    \begin{equation}
        \alpha = \frac{a_{0}^{2}}{4\pi r_{+}}
    \end{equation}
Next, consider a general term in the summation over $n$ in equation (\ref{tildeCseries}), which occurs at order $g^{n}$: $\langle \phi_{R} O_{L}^{n} O_{R}^{n} \phi_{L} \rangle$. As pointed out in \cite{Maldacena:2017axo}, at higher orders in $g$, the gravitational interaction continues to dominate over other kinds of interactions, and we can view the term above as consisting of $n$ independent scattering events. After multiplying the phases together and resumming the series over $n$, we find:
\begin{equation}
    \tilde{C} = \alpha \int dp^{U} dx p^{U} \psi_{2}^{*}(p^{U},x) \psi_{3}(p^{U},x) \mathrm{exp} \left[ i \alpha g \int dp^{V} dy p^{V} \psi_{1}^{*}(p^{V}, y) \psi_{4}(p^{V}, y) e^{i \delta(s)} \right]
    \end{equation}
We have thus derived equation (\ref{Ctilde1}). For the example of the planar BTZ black hole discussed in the main text, we also write down here the four wavefunctions $\Psi_{1}$, $\Psi_{2}$, $\Psi_{3}$ and $\Psi_{4}$:
\begin{equation}\label{psi1}
\Psi_{1}{(p^{V},x)} = \frac{r_{+}^{\DO}}{2} e^{-r_{+}t_{0}\DO} e^{-2i p^{V} \cosh{(r_{+}(x-x_{0}))} e^{-r_{+}t_{0}}} \theta{(p^{V})} ( p^{V})^{\DO-1} \frac{e^{i\pi\DO/2} }{\Gamma{(\DO)}}
\end{equation}
\begin{equation}\label{psi2}
\Psi_{2}{(p^{U},x)} = \frac{r_{+}^{\Dpsi}}{2} e^{-r_{+}t_{R}\Dpsi} e^{-2i p^{U} \cosh{(r_{+}(x-x_{R}))} e^{-r_{+}t_{R}}} \theta{(p^{U})} ( p^{U})^{\Dpsi-1} \frac{e^{i\pi\Dpsi/2} }{\Gamma{(\Dpsi)}}
\end{equation}
\begin{equation}\label{psi3}
\Psi_{3}{(p^{U},x)} = \frac{r_{+}^{\Dpsi}}{2} e^{r_{+}t_{L}\Dpsi} e^{2i p^{U} \cosh{(r_{+}(x-x_{L}))} e^{r_{+}t_{L}}} \theta{(p^{U})} ( p^{U})^{\Dpsi-1} \frac{e^{-i\pi\Dpsi/2} }{\Gamma{(\Dpsi)}} 
\end{equation}
\begin{equation}\label{psi4}
\Psi_{4}{(p^{V},x)} = \frac{r_{+}^{\DO}}{2} e^{r_{+}t_{0}\DO} e^{2i p^{V} \cosh{(r_{+}(x-x_{0}))} e^{r_{+}t_{0}}} \theta{(p^{V})} (p^{V})^{\DO-1} \frac{e^{-i\pi\DO/2} }{\Gamma{(\DO)}} 
\end{equation}

\subsection{From $\tilde{C}$ to $C$}
Having computed $\tilde{C}$ in the previous subsection, we need relate this to the quantity $C$, the imaginary part of which indicates communication through the wormhole.  Recalling that $C = e^{-i \langle V\rangle}\tilde{C}$, the two differ by an overall phase $\langle V \rangle= \frac{g}{K} \sum_{i=1}^{K} \langle \mathcal{O}_{R}^{i}(t_0,x_0) \mathcal{O}_{L}^{i}(t_0,x_0) \rangle= \frac{g}{K} \sum_{i=1}^{K} \langle O_{R}^{i} O_{L}^{i} \rangle$, where we adopt the same compact notation as the previous section. We focus pn one term in the summation over $i$. We think of this 2-point function as the overlap between 2 states, and in-state $|\psi \rangle$ and an out-state $|\psi' \rangle$:
\begin{equation}
    |\psi\rangle = O_{R} |TFD\rangle
\end{equation}
\begin{equation}
    |\psi'\rangle = O_{L} |TFD\rangle
\end{equation}
We can borrow the the two relevant wavefunctions from the previous subsection of this appendix. With the subscripts $1$ and $4$ representing $O_{L}$ and $O_{R}$ respectively we have
\begin{equation}
    |\psi\rangle = \int \Psi_{4} {(p^{V},x_{4})} | p^{V} x_{4} \rangle_{in}
\end{equation}
\begin{equation}
    |\psi' \rangle = \int \Psi_{1} {(p^{V},x_{1})}| p^{V} x_{1} \rangle_{out}
\end{equation}
and the 2-point function becomes:
\begin{equation}
    \langle \psi' | \psi \rangle = \int  \Psi_{1}^{*}{(q^{V},x_{1})} \Psi_{4}{(p^{V},x_{4})} {}_{out} \langle q^{V} x_{1} | p^{V} x_{4} \rangle_{in}
\end{equation}
This is simply proportional to (\ref{D}) in the main text, but lacking a phase shift (i.e. with $G_N=0$). Following through the integrals as in the main text simply results in $H(\Delta_O)$ of equation (\ref{eq:HDO}). So $C$ relates to $\tilde{C}$ through
\begin{equation}\begin{split}
        C=&\exp\left(-i g H(\DO)\right)\tilde{C}
\end{split}\end{equation}

\section{Bounding the information velocity}
\label{app:vE_bound}

In this appendix, we argue for the inequality
\begin{equation}
    \frac{v_E(f)}{1-f} \leq v_B.
\end{equation}
This is done by generalizing an argument due to Afkhami-Jeddi and Hartman~\cite{Hartman_2015_Bound_vE}. In that work, it was shown for Lorentz invariant field theories that $v_E(f) \leq c$, where $c$ sets the microcausal speed limit beyond which commutators identically vanish.

We propose to improve this bound in two ways. First, the microcausal speed limit $c$ should be replaced by an effective light-cone speed limit $v_L$. Later we will argue that $v_L = v_B$. Second, the explicit dependence on $f$ should be brought out. While Ref.~\cite{Hartman_2015_Bound_vE} does actually bound the general $v_E(f)$, the bound is quite loose near $f=1$. 

As in the main paper, we focus on strip regions in any dimension. The important quantity is the relative entropy between a non-equilibrium state of a given energy density and entanglement fraction, $\psi$, and the corresponding thermal equilibrium state, $\sigma \propto e^{-H/T}$,
\begin{equation}
    S_{\text{rel}}(A,t)=S(\psi_A(t)| \sigma_A).
\end{equation}
In the first part of this appendix, we sketch the argument in simplified form, ignoring various subtleties. In the other two parts, we argue for $v_L = v_B$ and clarify some technical details that are glossed over in the simplified argument.

\subsection{Simplified argument}

\begin{figure}
    \centering
    \includegraphics[width=.9\textwidth]{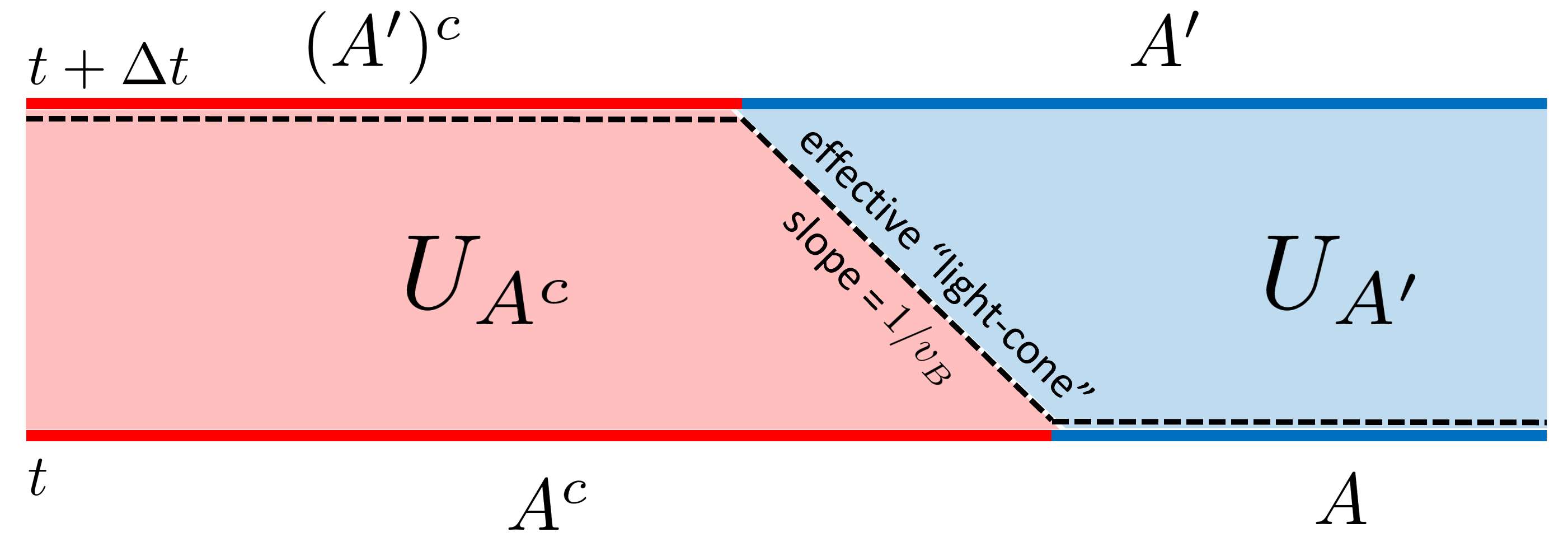}
    \caption{Schematic illustrating the decomposition of the time-evolution operator, $U \approx U_{A'} U_{A^c}$, along the effective light-cone set by the butterfly velocity $v_B$ (or by the light-cone speed $v_L$, when this is not equal to $v_B$).}
    \label{fig:circ_dec}
\end{figure}

Let us assume there is a light-cone velocity $v_L$, potentially depending on energy density $\epsilon$, such that information does not significantly propagate beyond the effective light-cone. Here is a precise version: Given a strip $A$, we can widen $A$ by an amount $v_L \Delta t$ on each side (see Figure~\ref{fig:circ_dec}) to produce a region $A'$ such that the time evolution $U$ can be decomposed as 
\begin{equation}
    U \approx U_{A'} U_{A^c}
\end{equation}
when acting on any translation invariant state of the given energy density. Here $U_{A^c}$ is supported entirely on the complement $A^c$ of $A$ and $U_{A'}$ is supported entirely on $A'$. We will specify the precise meaning of $\approx$ in the detailed argument below.

One implication of this statement is that, given access to region $A'$ at time $t+\Delta t$, the dynamics can be entirely reversed as far as $A$ is concerned. In this sense, any information that was initially in $A$ is definitely in $A'$ at later time since the state of $A$ can be recovered from the state of $A'$ by applying $U_{A'}^{-1}$.

With this decomposition, we have the following monotonicity result,
\begin{equation}
    S_{\text{rel}}(A',t+\Delta t) \geq S_{\text{rel}}(A,t),
\end{equation}
where, again, $A'$ is the widening of $A$ by $v_L \Delta t$. This is proven using the unitary invariance and monotonicity of the relative entropy.

Given the modular Hamiltonian $K_{ A}$ of $\sigma_A$, the relative entropy is
\begin{equation}
    S_{\text{rel}}(A,t)=S(\sigma_A)-S(\psi_A(t)) - \langle K_{A} \rangle_{\sigma} + \langle K_A \rangle_{\psi(t)}.
\end{equation}
Using the fact that the modular Hamiltonian of the thermal equilibrium state in region $A$ is the restriction of the Hamiltonian to $A$, up to edge effects, it follows that the modular Hamiltonian terms cancel in the relative entropy formula since $\psi_A$ and $\sigma_A$ have the same energy density. Hence, the relative entropy reduces to
\begin{equation}
    S_{\text{rel}}(A,t)=S(\sigma_A)-S(\psi_A(t)).
\end{equation}
Using the monotonicity result combined with this formula for the relative entropy gives
\begin{equation}
    S(\sigma_{A'}) - S(\psi_{A'}(t+\Delta t)) \geq S(\sigma_{A}) - S(\psi_{A}(t)).
\end{equation}

We now need to compute several entropy differences. The difference of thermal state entropies is
\begin{equation}
    S(\sigma_{A'}) - S(\sigma_{A}) = s (|A'| - |A|).
\end{equation}
Hence, we have 
\begin{equation}
    S(\psi_{A'}(t+\Delta t)) - S(\psi_{A}(t))  \leq s (|A'| - |A|).
\end{equation}
If the entropy of $A$ is not yet saturated, the specified entanglement fraction implies
\begin{equation}
    S(\psi_{A}(t)) = S(\psi_{A'}(t))- f s (|A'| - |A|).
\end{equation}
Using the definition of $v_E$, the difference of entanglement of $A'$ at time $t$ and $t+\Delta t$ is
\begin{equation}
    S(\psi_{A'}(t+\Delta t))-S(\psi_{A'}(t)) = 2 L^{d-1} s v_E(f)\Delta t,
\end{equation}
where $L$ is the length of the strip in the transverse directions. The volume difference can also be expressed in terms of the same ingredients,
\begin{equation}
    (\text{vol}_{A'} - \text{vol}_A) = 2 L^{d-1} v_L \Delta t.
\end{equation}

Combining these entropy relations for $A$ and $A'$ into the monotonicity relation gives
\begin{equation}
    2 L^{d-1} v_E(f) \Delta t \leq 2 L^{d-1} (1-f) s v_B \Delta t.
\end{equation}
Cancelling the factors of size and entropy, we finally have the result,
\begin{equation}
    v_E(f) \leq (1-f) v_L.
\end{equation}
Assuming that $v_L = v_B$ gives the desired result; this is argued for in the next subsection below.

\subsection{Light-cone velocity and butterfly velocity}

Here we argue that $v_L = v_B$. This is done by directly constructing the approximation $U \approx U_{A'} U_{A^c}$. The main idea is to break the Hamiltonian up into two pieces, one in $(A')^c$ and one in $A'$. Using the interaction picture, the evolution then breaks up into a piece strictly contained in $A'$ and a piece initially contained in $(A')^c$ which spreads with time. If the distance between $(A')^c$ and $A$ is big enough, then we can truncate the second part of the evolution to sit entirely inside $A^c$. This yields the desired form.

We use a lattice notation where the Hamiltonian is 
\begin{equation}
    H = \sum_r h_r.
\end{equation}
We have not made the effort to a fully continuum argument, so any field theory will have be viewed as lattice regulated. Let $H_{A'}$ denote the sum of all terms strictly contained in $A'$:
\begin{equation}
    H_{A'} = \sum_{r: h_r \in A'} h_r.
\end{equation}
The remainder is
\begin{equation}
    \check{H}=H-H_{A'}.
\end{equation}
Note that $\check{H}$ and $H_{A'}$ do not commute in general.

The interaction picture with respect to $H_{A'}$ is defined by the unitary $U_I$ obeying $U_I(t'=0)=1$ and
\begin{equation}
    i\frac{d U_I(t')}{dt'} = e^{i H_{A'} t'} \check{H} e^{-i H_{A'} t'} U_I,
\end{equation}
such that the total time evolution is
\begin{equation}
    U = e^{-i H_{A'}t} U_I(t'=t).
\end{equation}
The goal is to truncate $U_I$ so that it lies entirely inside $A^c$. To do so, we need to understand the size of $e^{i H_{A'} t} \check{H} e^{-i H_{A'} t}$.

At $t=0$, this operator lies entirely in $(A')^c$ which we take to be a distance $\ell$ from $A$. At a time $t$, the question is then how large does $\ell$ need to be such that $e^{i H_{A'} t} \check{H} e^{-i H_{A'} t}$ has little support in $A$? However, as we emphasized this is too imprecise: what we care about is not the total support of this operator, but its effective support acting on states of a given energy density. The idea is that the square commutator in the thermal state of the same energy density should measure the effective size of operators. Based on this, we above assumed that $\ell = v_B t$ sufficed. To make this estimate more precise, we should consider the effects of wavefront broadening.

Following Ref.~\cite{Xu_2018_Scrambling_Fluctuations}, we will assume that an initially local operator evolved for time interval $\Delta t$ ($ t \rightarrow \Delta t$ in the above formulas) at energy density can be localized $\epsilon$ well on an interval of radius 
\begin{equation} \label{eq:buffer_size}
   \ell_0 = v_B \Delta t + \left( \frac{v_B (v_B \Delta t)^p}{\lambda}\log \frac{C_0}{\epsilon} \right)^{\frac{1}{1+p}}
\end{equation}
for constants $p$, $\lambda$, and $C_0$. This formula is chosen so that the squared commutator of the time-evolved local operator with any operator more than $\ell_0$ away is less than $\epsilon$. This estimate is obtained assuming the squared commutator takes the form
\begin{equation}
    C(r,t) \sim C_0 \exp \left( - \lambda \frac{(r-v_B t)^{1+p}}{v_B (v_B t)^p}\right)
\end{equation}
when $r \gg v_B t$~\cite{Xu_2018_Scrambling_Fluctuations}.

Observe that $e^{i H_{A'} t} \check{H} e^{-i H_{A'} t}$ is a sum of local terms each of which is initially at least $\ell$ (the distance between $A$ and $(A')^c$) away from $A$. Using the above estimate for operator size, we suppose that, if $\ell > \ell_0$, then all the terms in $e^{i H_{A'} t} \check{H} e^{-i H_{A'} t}$ can be at least $\epsilon$ well localized on $A^c$. Indeed, there is a further suppression the further away from $A$ the terms originate. Hence, if $\ell > \ell_0$, we argue that $U_I$, when acting on a state of the given energy density, can be replaced by an operator supported on just $A^c$ to precision $\epsilon$. Hence, the full time evolution can be approximated as $U\approx U_{A'} U_{A^c} $ to precision $\epsilon$ assuming $\ell > \ell_0$.

While this result is of course non-rigorous, it captures our expectation that $v_B$ sets the effective speed limit. At infinite temperature, we believe the argument can be made rigorous using the trick of averaging over all operators to bound the deviation from the identity operator and using the bounds in Ref.~\cite{Xu_2018_Scrambling_Fluctuations} to bound commutators of complex operators in terms of those of simpler operators.

We also comment that we have assumed that $H_{A'}$ has the same butterfly speed as $H$, at least far from the cut separating $A'$ and $(A')^c$. In a similar vein, one could worry that these cuts are rather drastic and introduce high energy perturbations to the analysis. Both effects can be mitigated by smearing the terms in $H$ a little in time to restrict their off-diagonal matrix elements to a low energy shell and by making the transition from $A'$ to $(A')^c$ soft by including a slowly varying envelope function into the definition of $H_{A'}$.

\subsection{Some technical points}

While the above discussion captures the conceptual essence of the argument, which is a minor variation of that in Ref.~\cite{Hartman_2015_Bound_vE}, there are several technical details that are not properly addressed. We next turn to these detailed points to make a more careful argument.

The first key point, as discussed in the preceding subsection, is that the separation between $A$ and $(A')^c$ must be somewhat larger than $v_B t$ due to wavefront broadening in the squared commutator. As discussed above, to have $\epsilon$ accuracy in the decomposition of the time-evolution operator, we must take the separation between $A$ and $(A')^c$ to be $\ell_0$ from Eq.~\ref{eq:buffer_size}. We note that $\ell_0$ depends only logarithmically on $\epsilon$, so even if $\epsilon$ is taken to scale as $1/|A|$ (the size of $A$), the required value of $\ell_0$ does not change dramatically.

The idea as above is to use the monotonicity of relative entropy. We are directly interested in the change in entropy of $\psi$ after evolving with $U(\Delta t)$. However, the strategy is to use monotonicity of relative for a closely related state, and then bound the entropy differences later using the Fannes-Audenaert inequality. Throughout this subsection, $U$ will refer to the evolution operator for a time $\Delta t$; note that $\Delta t$ will not necessarily be small. The monotonicity statement we use is
\begin{equation}
    S( [U_{A'} U_{A^c} \psi U_{A^c}^\dagger U_{A'}^\dagger]_{A'} |[U_{A'} U_{A^c} \sigma U_{A^c}^\dagger U_{A'}^\dagger]_{A'})\geq S(\psi_A|\sigma_A). 
\end{equation}
This is proven from monotonicity after using unitary invariance to remove the $U_{A'}$ factors on the left hand side.

The relative entropy between two states $\rho$ and $\sigma$ is
\begin{equation}
    S(\rho|\sigma)=-S(\rho)-\text{tr}(\rho \ln \sigma),
\end{equation}
so we must specify the modular Hamiltonian of $\sigma$ to compute the relative entropy. In our case, $\sigma$ is a thermal state or the restriction of a thermal state to a subsystem. We will assume that the modular Hamiltonian of such a state is
\begin{equation}
    \log \sigma_A = - \beta (H_A + \delta H_A) + \text{const}
\end{equation}
where $\beta=1/T$, $H_A$ is the physical Hamiltonian restricted to $A$, and $\delta H_A$ is a correction term localized near the boundary of $A$.

Hence, the relative entropy of state $\psi(t)$ with respect to $\sigma$ is
\begin{equation}
    S(\psi_A(t)|\sigma_A) = S(\sigma_A)-S(\psi_A(t)) + \langle \beta (H_A + \delta H_A)\rangle_{\psi(t)} - \langle \beta (H_A + \delta H_A)\rangle_{\sigma}.
\end{equation}
However, the monotonicity statement above connected relative entropies for states evolved with the approximate evolution. We must also specify how the modular Hamiltonian of $[U_{A'} U_{A^c} \sigma U_{A^c}^\dagger U_{A'}^\dagger]_A$ looks. Since $\sigma$ is exactly preserved by the true evolution, we assume that the approximate evolution approximately preserves $\sigma$ (to $\epsilon$ accuracy). 

In particular, modifications to $\sigma_A$ should be localized near the edge of $A$. Hence, combined with the effects of the correction term $\delta H_A$, we have
\begin{align}
    & S( [U_{A'} U_{A^c} \psi U_{A^c}^\dagger U_{A'}^\dagger]_{A'} |[U_{A'} U_{A^c} \sigma U_{A^c}^\dagger U_{A'}^\dagger]_{A'}) = \nonumber \\
    &\,\,\,\, s |A'| - S(\psi_{A'}(t+\Delta t))+  O(|\partial A'|) + O(\epsilon |A'|).
\end{align}
The first two terms appeared in the ideal argument and are what we want, while the second two reflect corrections due to uncontrolled edge effects (both from $\delta H_A$ and the approximate time evolution) and possible changes in entropy between the exact evolvted state $\psi_A(t)$ and the approximately evolved state.

With the above formula, monotonicity implies
\begin{align}
  s(|A'|-|A|) +O(|\partial A'|) + O(\epsilon |A'|) \geq S(\psi_{A'}(t+\Delta t )) - S(\psi_{A}(t)).
\end{align}
Taking $\epsilon \ll 1/|A'|$ suffices to make the $O(\epsilon |A'|)$ term small without causing a significant blow-up in the value of $\ell_0$. The $O(|\partial A'|)$ term can be made to give a relatively small contribution by making $\Delta t$ large. 

Specifically, if the difference in volumes is 
\begin{equation}
    |A'|-|A| = 2 L^{d-1} \ell_0,
\end{equation}
and if, by assumption, the difference of $\psi$ entropies is
\begin{equation}
    S(\psi_{A'}(t+\Delta t )) - S(\psi_{A}(t)) = 2 L^{d-1} s v_E(f) \Delta t +f s (|A'|-|A|),
\end{equation}
then it follows that
\begin{equation}
    2 L^{d-1} s (1-f) \ell_0 + O(|\partial A'|) \geq 2 L^{d-1} s v_E(f) \Delta t.
\end{equation}
Using the expression for $\ell_0$ (Eq.~\ref{eq:buffer_size}) and dividing through by $ 2 s L^{d-1} \Delta t$ gives
\begin{equation} \label{eq:vE_bound_corrections}
    (1-f) v_B + O\left( \frac{1}{ \Delta t}\left( \frac{v_B (v_B \Delta t)^p}{\lambda}\log \frac{C_0}{\epsilon} \right)^{\frac{1}{1+p}} , \frac{|\partial A'|}{s L^{d-1} \Delta t }\right) \geq v_E.
\end{equation}

The crucial final claim is that we can choose $|A|$ and $\Delta t$ large enough such that the linear entanglement growth assumption is valid and the correction terms in Eq.~\eqref{eq:vE_bound_corrections} are small. Since $|\partial A'| \sim L^{d-1}$ (and since we argued that the length scale of these corrections was bounded), the only dangerous term is the first one. As long as $p/(1+p) < 1$, which is $p>0$, these conditions can be met by choosing \begin{equation}
    \text{width of $|A|$} \gg v_E \Delta t \gg 
   \frac{v_B}{\lambda} \log \frac{C_0}{\epsilon}.
\end{equation}
The first inequality guarantees that we are still in the linear growth regime and the second inequality guarantees that the correction term is small. Since we may take $\epsilon \sim 1/|A'|$, these conditions are simultaneously achievable since they are roughly of the form $\text{width} \gg \Delta t \gg \log \text{width}$.

Hence, we conclude that 
\begin{equation}
    (1-f) v_B \geq v_E(f) - \delta
\end{equation}
for any $\delta >0$. More generally, if the light-cone speed is $v_L$ with $v_L \neq v_B$, we learn that
\begin{equation}
    (1-f) v_L \geq v_E(f) - \delta.
\end{equation}

\bibliography{bib/bibliography.bib, bib/scrambling.bib, bib/mis.bib}

\end{document}